\documentclass[a4paper,notitlepage,12pt]{article}
\usepackage{amssymb}

\usepackage{amsmath}
\usepackage{epsfig}
\usepackage{graphicx}

\begin{document}

\title{\textbf{Endophysical information transfer in quantum processes}}
\author{\textbf{Jon Eakins and George Jaroszkiewicz} \\
School of Mathematical Sciences, University of Nottingham,\\
University Park, Nottingham NG7 2RD, UK}
\date{\today}
\maketitle

\begin{abstract}
We give a mathematical criterion for the concept of information flow within
closed quantum systems described by quantum registers. We define the
concepts of separations and entanglements over quantum registers and use
them with the quantum zip properties of inner products over quantum
registers to establish the concept of partition change, which is fundamental
to our criterion of endophysical information exchange within such quantum
systems.

\vspace{0.25in}

\noindent PACS number(s): 03.65.Ca, 03.65.Ta, 03.65.Ud, 03.67.Lx
\end{abstract}

\newpage

\begin{center}
{\large {$\mathbf{I.}$ \textbf{INTRODUCTION}} }
\end{center}

This paper aims to give a mathematical definition of information exchange
within quantum systems. This is motivated by the still unresolved questions
of what an observer is and the meaning of measurement in quantum physics.
Collectively, these and related issues will be called the \emph{measurement
problem}.

Before we discuss the definition of information exchange, we should say what
we mean by information. According to Preskill \cite{PRESKILL-98},
information is something that is encoded in a state of a physical system.
According to Sippl \cite{SIPPL:80}, however, information is knowledge that
was not previously known to its receiver. These two definitions do not seem
equivalent to us. In fact, they are very different. The former presents a
passive, classical perspective which suggests that information is a property
of a system which is ``there'' waiting to be found, regardless of anything
else in the universe. On the contrary, the latter definition is thoroughly
dynamic, requiring both subject and observer and the passage of time for the
definition to make any sense. We will discuss information according to the
Sippl point of view, because it accords better with quantum principles.
Quantum theory is not about systems. It is about interactions between
systems and is therefore all about time. Data held in a system has no
physical meaning in the absence of any measurement of that data. In this
article, therefore, the terms \emph{information}, \emph{information
exchange, information acquisition and loss} will be regarded as synonymous.
The term ``data'' will be used to refer to mathematical properties encoded
into state vectors representing physical systems.

\

Quantum mechanics is such an accurate and practical tool in the physical
sciences that we should explain our interest in the measurement problem. In
addition to a natural interest in constructing a completely consistent
quantum theory, we have been motivated by two long-standing problems of
modern theoretical physics; first, quantum field theory is littered with
mathematical divergences and second, quantum gravity, the program attempting
to rewrite Einstein's classical theory of general relativity in quantum
terms, has had limited success. It seems to us that the problems with
quantum field theory and quantum gravity lie not with the general principles
of quantum mechanics but with three historical legacies related to the
measurement problem that these theories acquired at birth.

The first of these is the supposed continuity of space and time. Both
quantum field theory and classical general relativity assume that space and
time form a continuum in which various sorts of field can be embedded. Many
attempts at quantum gravity also implicitly suppose the existence of some
underlying manifold with a fixed dimension. The same is true of string
theory and developments of it such as brane physics. Our view is that
classical spacetime is a throwback to the prequantum, classical mechanical
view of the world sometimes referred to as the block universe \cite{PRICE:97}%
. There is no logical reason to suppose that any concept which originated
before the advent of quantum mechanics should survive as an intrinsic one in
a fully quantum theory of the universe. The many attempts to solve the
problems associated with continuity rely on the introduction of ad hoc
modifications to the continuum, such as spacetime lattice discretization,
point splitting and dimensional regularization, all of which simply
reinforce the view that spacetime continuity is a terminally sick concept.

The supposed continuity of space and time cannot in fact be empirically
proved. Indeed, rather like pre-atomic continuum theories of liquids, it is
an abstraction which arises from the particular way in which humans interact
with their environment. Normally, this environment is so flooded with
photons that the brain can maintain a consistent illusion that we exist in a
three-dimensional continuum and evolve according to a continuous time. This
gives a classical picture of a universe which advanced technology shows runs
on quite different quantum principles.

The second historical legacy weighing down modern theory has been recognized
and taken seriously by physicists only relatively recently. This is the
issue of \emph{exophysics} versus \emph{endophysics}. In exophysics, the
assumption is made that observers stand outside of the systems that they
observe. This perspective is the basis of the standard Copenhagen School
approach to quantum mechanics. In endophysics, on the other hand, observers
and systems are all part of a greater whole. The ultimate expression of the
endophysical perspective is the statement that there is only one system, the
universe, which contains absolutely everything, including all forms of
observers.

One essential difference between exophysics and endophysics lies in the
meaning of information acquisition and storage. According to the exophysical
perspective, whenever an observer measures something about a system,
information is registered in some form of memory carried by that observer.
The exophysical principles of classical mechanics generally assume that this
registration process can be done without affecting the system being observed
and effects the observer only via changes in the memory. This idea arises
naturally, given the classical, three-dimensional model of external reality
constructed by the human brain. The nature of this memory is rarely, if
ever, discussed in exophysics, whereas it becomes crucial in endophysics to
explain in what sense endophysical observers record ``measurements''.

When applied to quantum physics, the exophysical perspective creates its own
problems. First, quantum mechanics has been found to be valid at all scales
looked at, so that no natural ``Heisenberg cut'' (the hypothetical dividing
line between classical observers and quantum systems) seems to exist.
Second, quantum correlations seem oblivious to some of the properties
associated with classical relativistic spacetime, such as the principle of
local causes (i.e., Einstein locality \cite{PERES:93}). For example, quantum
correlation speeds vastly in excess of the speed of light have been reported
recently \cite{SCARANI-00}.

Changing to the endophysical perspective appears to remove the problem with
the Heisenberg cut, because observers are now regarded as quantum subsystems
within a larger quantum system, so that quantum principles can cover
everything without the need for any sort of cut. Unfortunately, changing
perspective merely replaces one problem with another: now we have to explain
what the difference between an observer and a system under observation is
and what an act of measurement means.

Another consequence of this change is that there arises the extraordinarily
difficult task of explaining how the classical world that we see on
macroscopic scales could arise from a purely quantum theory. This program
will be referred to as the problem of \emph{emergence}. If we went further
and assumed no a priori Riemannian geometrical structures whatsoever,
anticipating their appearance only in some emergent limit, then we would be
dealing with what Wheeler has called \emph{pregeometry} \cite{WHEELER-80}.

The third historical legacy inherited by modern physics is its unwillingness
to completely let go of all classical modes of thought, particularly
concerning the concept of \emph{observer}. Relativistic covariance
principles tend to be applied at all levels of modern theory, whereas more
careful analysis reveals that these principles cannot be upheld everywhere
at all costs. The problem occurs because the classical relativist believes
that different observers can observe the same event, but the quantum
theorist knows that this is physically incorrect. In spite of this, the
language and thinking of relativity is predicated on the former point of
view, which continues to infect modern theory at all levels. We shall
discuss this issue in the context of the process time perspective in the
next section.

\

In earlier work \cite{JAROSZKIEWICZ-03B} we explored the idea discussed by
Feynman \cite{FEYNMAN-82} and others that the universe can be represented in
terms of a vast \emph{quantum register}, i.e., a tensor product of a large
number of quantum subregisters such as qubits. In our approach, the dynamics
of the universe is postulated to be that of a self-referential quantum
register with no external observers.

This approach has a number of important and useful properties. First, it is
based on Hilbert space rather than on classical configuration or phase
space. Therefore, rather than trying to quantize a classical theory in the
traditional way, the quantum register approach starts off completely
consistent with quantum principles. Second, the qubits making up the full
quantum register provide the ultimate source of the vast number of degrees
of freedom which the physical universe is known to have (the reader is
warned that the relationship between these concepts is not one-to-one and is
considerably more subtle than expected). Third, we have shown that the
properties of factorization and entanglement associated with quantum
registers can be used to describe causal set dynamics in a natural way \cite
{JAROSZKIEWICZ-03B}. We found that the elements of the causal sets involved
are not in fact the subregisters as might be expected. It is the subtle and
intricate interplay between the patterns of factorization and entanglement
in both the states of the quantum register and the Hermitian operators over
that register which generates the causal set dynamics.

A particular feature of our approach is its use of state reduction rather
than the unitary, no-collapse evolution favoured in the many-worlds and
decoherence paradigms. There are several reasons for this. First, we wish to
discuss physics as it appears to the experimentalist; state reduction
corresponds to the registration of information in quantum experiments
whereas Schr\"{o}dinger evolution holds in the absence of such registration.
Second, although the standard view is that quantum mechanics predicts only
expectation values, it is an empirical fact that interference patterns can
be built up over vast time scales from a succession of single outcomes which
are well separated in time. This and the extraordinary difficulties
encountered by hidden variables theories to account for all current
experimental data suggests that single quantum outcomes do have an
individual physical significance, albeit not a classical one. Probability is
not just about averages and expectation values. It is too easily turned into
a mathematical set-theoretic discussion of ``measure'', whereas in practice
it is bound up with the counting of frequencies of quantum outcomes by
observers.

A third reason is that the Schr\"{o}dinger evolution of quantum systems can
always be unitarily transformed away by moving to the Heisenberg picture,
supporting the view that temporal evolution is not an intrinsic property of
\emph{systems} on their own, but of those systems and of the \emph{observers}
of those systems as well. Another way of saying this is that physical time
is a marker of quantum information exchange.

\

The structure of this paper is as follows. In $\S 2$ we review the stages
paradigm, which is the conceptual framework in which we work. In $\S 3$ we
discuss its extension to quantum registers. In $\S 4$ we discuss the
separation and entanglement properties of quantum registers, introducing the
important concept of a lattice of \emph{partitions}, on which we base our
ideas of endophysical information exchange within quantum systems. In $\S
5,6 $ and $7$ we discuss operators, eigenvalues, preferred bases and the
separability properties of such bases. In $\S 8$ we discuss the relationship
between active and passive transformations, as this is a central issue in
quantum dynamics, followed in $\S 9$ by the concept of local
transformations, which are relevant to quantum registers. In $\S 10$ we
discuss what is meant by state preparation, followed in $\S 11$ by a
discussion of transition amplitude factors. In $\S 12$ and $13$ we focus
attention on the concept of an isolated quantum system. In $\S 14$ we state
our principle of endophysical information exchange, based on the concept of
partition change, followed in $\S 15$ by some concluding remarks.

In this paper we shall represent state vectors in two way; when we discuss
more abstract issues such as quantum cosmology we favour symbols such as $%
\Psi $, but when we have more detailed statements to make we use the
equivalent notation $|\Psi \rangle $. This latter notation is more useful in
the representation of operators. Inner products will also be represented in
two ways, i.e., we will take
\begin{equation}
\left( \Psi ,\Phi \right) =\langle \Psi |\Phi \rangle .
\end{equation}

\

\begin{center}
{\large {$\mathbf{II.}$ \textbf{THE STAGES PARADIGM}} }
\end{center}

Our account of time and information exchange rests on two observations: $i)$
the laws of quantum mechanics appear to have universal application and $ii)$
the universe contains a truly vast number $N$ of degrees of freedom.
Throughout our work we shall assume $N$ is finite, principally because this
ensures that we do not have any divergences in any of our equations and that
all our Hilbert spaces are separable. In earlier work we used both of these
ideas to develop further Feynman's view of the universe as a gigantic form
of quantum computation \cite{FEYNMAN-82} which behaves as an autonomous
quantum system with no external observers \cite
{JAROSZKIEWICZ-01A,JAROSZKIEWICZ-02A}. We shall refer to this as the \emph{%
stages paradigm}, the essential details of which are reviewed briefly as
follows.

Because the laws governing observers of quantum systems are currently not
understood \cite{FEYNMAN-82}, the conventional Copenhagen School approach to
quantum mechanics assumes observers to be semiclassical objects with free
will, standing outside of the quantum systems they are investigating. It is
therefore an exophysical approach. In mathematical terms this translates to
the use of differential equations governed by boundary conditions dictated
by arbitrary factors, such as choice of initial state and experiment,
external to the quantum systems under discussion.

Our approach is different. We take it as self evident that the universe
itself organizes its own observations (or tests \cite{PERES:93}), because by
definition the universe contains everything and therefore there can be no
semiclassical observers standing outside it. This forces us to make one (and
only one) change in the standard principles of quantum mechanics: we have to
remove the rule that there are semiclassical observers external to quantum
systems deciding how to measure them.

By this we do not mean that the concept of a semiclassical observer is
wrong. Of course, it has proved to be enormously useful. What we do mean is
that this notion is not an intrinsic or essential one. We believe that it is
a derived or emergent aspect of quantum registers when looked at on certain
scales, typically when extremely large numbers of subregisters are involved
and when the complex factorization and entanglement properties of quantum
registers permit it. In this paper we discuss what the notion of
semiclassical observer should be replaced by.

In common with all theories, there are some aspects of our formalism which
are unphysical, being auxiliary mathematical devices required to represent
the physics. The quantum register itself and its states are examples of such
devices. Another one is our concept of \emph{exotime; }this is the
underlying discrete time parameter, indexed by the integers, which labels
successive states of the universe. This time is not an observable. How it
relates to the time seen by endophysical observers (\emph{endotime}) is a
problem analogous to the question of how co-ordinate time relates to proper
time in relativity. The former is integrable (i.e., path independent) and
unphysical, whereas the latter is not and has direct physical significance.

At each instance $n$ of exotime the universe is assumed to be in a well
defined \emph{stage} $\Omega _{n}$. A given stage $\Omega _{n}\equiv \Omega
\left( \Psi _{n},I_{n},\mathcal{R}_{n}\right) $ consists of three things:

\begin{enumerate}
\item[i)]  a pure state $\Psi _{n},$ known as the \emph{state of the universe%
}, being a normalized element in a universal Hilbert space $\mathcal{H}$ of
some extremely large but finite dimension $N$. The state of the universe $%
\Psi _{n}$ is an eigenstate of some test $\Sigma _{n}$, i.e., $\Psi _{n}$
represents an outcome of some immediately previous test of the universe;

\item[ii)]  the \emph{current information content} $I_{n}$, representing
dynamical information over and above that contained within $\Psi _{n}$. For
example, the dynamical laws governing future states of the universe may
require a knowledge of which particular test $\Sigma _{n}$ (or equivalently,
which \emph{preferred basis}) gave $\Psi _{n}$ as an outcome;

\item[iii)]  the \emph{current rules} $\mathcal{R}_{n}$ (or laws of
physics), which determine how the current stage evolves.
\end{enumerate}

These ingredients represent the minimum we believe is needed to model any
self-referential quantum universe. Indeed, the equivalent of these
ingredients are implicit in the alternative many-worlds and decoherence
approaches; they assume states of the universe, Hamiltonian operators
(equivalent to dynamical information about the system), and the
Schr\"{o}dinger equation, which gives the rules for dynamical evolution.

\

\begin{center}
\textbf{A. Stage dynamics}
\end{center}

An essential feature of the stages paradigm is the dynamical relationship
between successive stages, i.e., the jump from $\Psi _{n}$ to $\Psi _{n+1}$.
Generalizing the standard rules of quantum mechanics \cite{PERES:93}, we
assume that each given stage $\Omega _{n}$ is replaced by another one, $%
\Omega _{n+1}$, such that the latter's state of the universe $\Psi _{n+1}$
is a quantum outcome (an eigenstate) of some quantum test $\Sigma _{n+1}, $
which itself was determined in some way by $\Omega _{n}$ and by no other
factor. One way of seeing stage dynamics is as a perpetual sequence of ideal
measurements \cite{PERES:93} interlaced with a sequence of self-determined
tests.

In the real world of the human observer, the physical experience of time
appears very different to its representation in classical theories such as
Newtonian mechanics and relativity. Real time appears to us not as a
continuum but as a single point, the enigmatic ''moment of the now'' also
known as the \emph{present}, albeit it is one which appears to change
constantly. Only our memories and our ability to anticipate the future allow
us to think of the past and the future, all of which thinking takes place in
the present. This view of time has been labelled \emph{process time}, in
contrast to the \emph{manifold time }or block universe perspective used
widely throughout relativistic physics, which is a geometric view of space
and time including past, present and future.

It is here that a conceptual clash occurs between the principles of quantum
mechanics and the principles of relativity. The Kochen-Specker theorem \cite
{KOCHEN+SPECKER-67} states that quantum systems do not have classical
properties per se waiting to be discovered. Taking this to its logical
conclusion, quantum mechanics has to deny the reality of the future and
therefore should strictly avoid using block universe concepts. The logic
behind this is based on the following assertions: $i)$ it is self-evident
that the future cannot be more physically real than the present and $ii)$
according to the Kochen-Specker theorem, the present is not there in any
classical sense. Therefore, the future is not there in any classical sense
and so the block universe model is fundamentally incorrect. This argument
also applies to the concept of closed timelike curves (CTCs) in general
relativity. These cannot be reconciled with the principles of quantum
mechanics, which is probably why they have never been observed.

There are two reasons why the block universe model is used throughout
theoretical physics. First, it allows us to order data in a way consistent
with the classical principles of Newtonian mechanics. These were formulated
from an exophysical perspective long before the discovery of quantum
mechanics and make use of an exophysical time concept known to Newton as
absolute time. Second, the concept of process time is not a Lorentz
covariant one, because simultaneity is frame dependent.

Relativity poses perhaps the most serious problem for the stages paradigm
from a number of perspectives. Apart from the problem with simultaneity, the
dimension of spacetime and the Lorentzian signature of the metric would have
to be explicable in terms of our pregeometric framework.

Actually, the mathematical structures associated with the stages paradigm
have particular properties which we believe can explain the emergence of
relativity. First of all, the issue of simultaneity is not the intractable
problem it appears to be. It arises because relativity itself arose from
purely classical ways of thinking about systems, observers, and what is
meant by observation. Consequently, some of the assertions frequently made
in the subject are either incorrect, incomplete, or incompatible with the
principles of quantum mechanics. For instance, there are actually no
infinitely extended inertial frames, yet many discussions in relativity
assume that they exist (we have in mind here the standard discussion of
elementary particle scattering in relativistic quantum field theory).

Another example is the widespread use of covariance arguments in relativity,
one of the most powerful principles employed in relativistic physics being
that the theory should be Lorentz covariant. This is misleading and
incorrect from a quantum measurement point of view. No single experiment is
Lorentz covariant. It sits in its own rest frame. Moreover, no single
outcome of any quantum experiment (i.e., a single run) can be ``observed''
by different observers sitting in different inertial frames. Otherwise we
could get around the Heisenberg uncertainty principle by having different
observers test a state for different incompatible observables. As emphasized
by Peres \cite{PERES-00B}, Lorentz covariance relates only to the
transformation properties of \emph{ensemble averages}, i.e., expectation
values, which is quite a different matter to what actually happens in any
single run of an experiment. Expectation values are statistical summaries of
information either already taken or planned to be taken from very many runs
of a given experiment. The stages paradigm on the other hand is designed to
discuss the process physics description of single runs, which has to take
place before ensemble averaging can be considered.

As for the other issues with relativity, we have shown \cite
{JAROSZKIEWICZ-03B} that a causal set structure arises naturally within the
stages paradigm once we extend it to large rank quantum registers. This then
opens the door to discussions of how concepts of space and Lorentz signature
metric can arise in emergent limits \cite{SORKIN+al-87}. Moreover, the
possibility of null tests \cite{JAROSZKIEWICZ-03B} permits the emergence of
a non-integrable endophysical time local to observers, thereby providing the
basis for the proper time concept in relativity. From the stages paradigm
point of view, therefore, relativity is but an emergent view of the quantum
universe and will not be discussed further here.

In the stages paradigm, the concept of process time cannot be modelled
directly; it is taken account of by the following rule: in any discussion,
only one stage (referred to as the \emph{present}) can ever be regarded as
certain. All other stages can be discussed only in conditional probability
terms relative to that stage. This reflects the essential feature of process
time, that only the ``present'' exists; the past is gone and the future has
no direct physical significance.

\

Each test $\Sigma _{n}$ is represented mathematically by some element $\hat{%
\Sigma}_{n}$ in $\mathbb{H}\left( \mathcal{H}\right) $, the set of all
nondegenerate Hermitian operators on $\mathcal{H}$. Nondegeneracy of
outcomes is as necessary here as in standard quantum mechanics, because
otherwise the interpretational power of quantum mechanics (such as the Born
probability rule) collapses. The eigenstates of $\hat{\Sigma}_{n}$
collectively constitute a unique (up to inessential phase factors of its
elements) basis $\mathsf{B}_{n}\equiv \mathsf{B}_{\Sigma _{n}}\left(\mathcal{%
H}\right) $ for $\mathcal{H}$, which gives a ``preferred basis'' at each
instant of exotime.

In standard quantum mechanics, attention is generally focused on
observables, their eigenstates and the corresponding eigenvalues, the theory
being particularly good at predicting the latter. The really important
problem physically, however, is how preferred basis sets should arise in the
first place. Eigenvalues in themselves have only a relative value, because
classical information, when it is expressed solely in the form of
eigenvalues, is not in general absolute. For instance, energy, charge and
momentum are all expressed relative to arbitrary levels and scales of
definition, which can only be done on emergent scales anyway. Moreover,
mathematically it is possible for different tests to correspond to the same
preferred basis set.

The stages paradigm incorporates these comments directly. There are no
Hamiltonians, no phase space, no differential equations of motion, and
eigenvalues are not taken to be important \emph{per se}. Instead, it is the
uniqueness of the elements of the preferred basis set which really matters.

\

The problem of how a preferred basis arises at each instant of exotime is an
unsolved problem, common to our stages paradigm, the many-worlds paradigm
and decoherence theory generally. If we knew the answer we would have a more
complete understanding of time and the universe. Whilst there is currently
no general understanding of how such bases arise in any theory, we shall
rule out free will in any shape or form and follow Feynman \cite{FEYNMAN-82}
in asserting that observers are part of the universe and are therefore
governed by its laws, whatever they are. The stages paradigm asserts that $%
\mathsf{B}_{n+1}$\ is determined solely by $\Omega _{n}$, i.e.,
\begin{equation}
\mathsf{B}_{n+1}=\mathsf{B}\left( \Omega _{n}\right) =\mathsf{B}\left( \Psi
_{n},I_{n},\mathcal{R}_{n}\right) .
\end{equation}

We do not exclude here the possibility that $\mathsf{B}_{n+1}$\ is itself a
random outcome of some higher form of quantum process, involving the
selection of one out of various potential elements of $\frak{B}\left(
\mathcal{H}\right) $, the set of all orthonormal bases for $\mathcal{H}$. If
there were some sort of random process governed by Born-type rules, this
would most appropriately be referred to as ``second quantization''. We
cannot rule out the possibility that the information content $I_{n}$
includes a knowledge of $\mathsf{B}_{n}$ and possibly of earlier preferred
bases, all of which could be used in the determination of $\mathsf{B}_{n+1}$.

Whatever the actuality, the stages paradigm is designed to describe our
existence in a well defined branch of reality and not in any superpositions
(as per many-worlds paradigm). We will assume that, given $\Omega _{n},$
there is always some subsequent selection process which picks out a definite
preferred basis $\mathsf{B}_{n+1}$ from $\frak{B}\left( \mathcal{H}\right) $.

\

\begin{center}
\textbf{B. Probabilities}
\end{center}

The stages paradigm accepts quantum randomness as an intrinsic property of
the universe which is quantified by the Born probability interpretation of
state vectors. Relative to a given stage $\Omega _{n}$ and to a given
preferred basis $\mathsf{B}_{n+1}$, the conditional probability (or
propensity) $P\left( \Psi _{n+1}=\theta ^{\alpha }\in \mathsf{B}%
_{n+1}|\Omega _{n},\mathsf{B}_{n+1}\right) $ of $\Psi _{n+1}$ being an
element $\theta ^{\alpha }$ of $\mathsf{B}_{n+1}$ is given by the Born rule
\begin{equation}
P\left( \Psi _{n+1}=\theta ^{\alpha }\in \mathsf{B}_{n+1}|\Omega _{n},%
\mathsf{B}_{n+1}\right) =|(\theta ^{\alpha },\Psi _{n})|^{2}.  \label{456}
\end{equation}
All basis states are taken to be normalized to unity, so we may write
\begin{equation}
\sum_{\alpha =1}^{\dim \mathcal{H}}P\left( \Psi _{n+1}=\theta ^{\alpha }\in
\mathsf{B}_{n+1}|\Omega _{n},\mathsf{B}_{n+1}\right) =1.
\end{equation}
Note that the probability (\ref{456}) is not in general the same thing as
the answer to the question: what is the probability of jumping to an
arbitrary element $\Theta $ in $\mathcal{H}$, given $\Psi _{n}?$ For most
elements in $\mathcal{H}$ the answer is \emph{zero}, even for those states $%
\Theta $ such that $|(\Theta ,\Psi _{n})|>0$. The reason is that $\Theta $
has to be an element of the preferred basis $\Sigma _{n+1}$ before there is
any possibility of such a jump occurring. In any such discussion, it is
important to keep in mind that tests are as important as states in
determining how the universe evolves. For instance, the microscopic
reversibility implied by the mathematical symmetry
\begin{equation}
|(\theta ^{\alpha },\Psi _{n})|^{2}=|(\Psi _{n},\theta ^{\alpha })|^{2}
\end{equation}
in (\ref{456}) should not be used in any discussion of physics without
careful consideration of the direction of time, the dynamical processes
concerned and the observers and environment involved.

The meaning of the probability rule (\ref{456}) when applied to the universe
requires careful interpretation, because it is here that criticisms of the
quantum universe concept have been raised \cite{FINK+LESCHKE-00}. By
definition the universe is not in an ensemble, so there is no direct
exophysical meaning to the concept of the probability of the outcome $\Psi
_{n+1}$.

We make two comments about this issue. First, in probability theory,
particularly in the Bayesian approach to statistics, there are two forms of
probability: \emph{epistemic uncertainty} is uncertainty arising from a lack
of knowledge whereas \emph{aleatory uncertainty} is uncertainty due to
inherent randomness. The former is reducible and may even be eliminated by
the acquisition of sufficient data. In quantum mechanics, this form of
uncertainty is encoded into the classical probabilities associated with
mixed states and is clearly predicated on the concept of some observer
external to a system having a lack of information about that system.
Aleatory uncertainty is irreducible and intrinsic, on the other hand. The
stages paradigm is based on the belief that quantum uncertainties are
inherently aleatory in nature. The problem with many-worlds and decoherence
is that they purport to derive quantum aleatory uncertainties from epistemic
foundations, which is inherently impossible. This accounts for their general
failure to explain the Born probability rule without the introduction of
extra assumptions which in the long run will amount to the state reduction
concept. It is for this reason that we have chosen to take state reduction
as a fundamental physical phenomenon in the first place.

The other point is that although cosmologists and quantum theorists are
themselves embedded in the universe and are therefore endophysical objects
in their own right, this does not prevent them from discussing possible
future states of the universe, including their own futures. This is somewhat
surprising given that another indirect criticism of quantum cosmology has
been the suggestion that G\"{o}del-type incompleteness rules out the
possibility of endophysical observers determining the exact state of the
universe they find themselves in \cite{BREUER-95}. While this may be true in
detail, this does not rule out cosmologists discussing \emph{conditional} or
relative probabilities, or making counterfactual statements about the
universe that they are part of.

\

\begin{center}
{\large {$\mathbf{III.}$ \textbf{QUANTUM REGISTERS}}}
\end{center}

Although the stages paradigm provides the basic framework for our quantum
dynamics, it is not specific enough as it stands to permit a discussion of
the measurement problem. We need to add to it the specific assumption that $%
\mathcal{H}$ is a quantum register consisting of a vast number $N$ of
subregisters.

By definition, a quantum register is a Hilbert space which is the tensor
product of a finite number of quantum subregisters, each of which is a
distinct Hilbert space of finite dimension. A simple calculation based on
Planck scales and the expansion of the observed universe suggests that $N$
has to be at least of the order $10^{180}$ if we want some chance of
describing the universe that we can see \cite{JAROSZKIEWICZ-02A}. It is
almost certainly much greater than that estimate, given that physical space
and all the currently known quantum fields describing matter should emerge
from such a pregeometric foundation. The dimension of $\mathcal{H}$ is at
least $2^{N}$, this lower bound occurring in the case that each subregister
is a qubit. This means that we are faced with the prospect of dealing with
mathematical structures capable of very great complexity indeed, which is a
double-edged sword. Whilst the mathematical structures associated with
quantum registers should in principle be capable of describing any physical
situation, their great complexity makes it near impossible to make detailed
calculation in any but the simplest situations.

In this article the separation and entanglement properties of states and
operators over quantum registers are crucial for the formulation of our
concept of endophysical information exchange. We shall focus our attention
on the relation between pure quantum states of a quantum register and \emph{%
strong} operators, which are a particular class of Hermitian operator acting
on those states and are discussed below.

It is frequently asserted that one of the crucial features of quantum
mechanics distinguishing it from classical mechanics is the occurrence of
entangled states. Whilst this is an important point, we have found that the
separable states are equally important, being used to represent classically
distinct observers and subsystems under observation. They represent the
nearest thing to classicality in quantum mechanics, because factors in a
separable state have an identifiable physical identity, something which
components in an entanglement do not have in the absence of measurement.
Separable states, therefore, are an essential component of our account of
endophysical measurement theory.

In the next section we introduce a notation designed to represent the
concepts of separations and entanglements of quantum registers.

\

\begin{center}
{\large {$\mathbf{IV.}$ \textbf{SPLITS, PARTITIONS, SEPARATIONS AND
ENTANGLEMENTS}}}
\end{center}

We shall use the notation $F\left[ x\right] \in \mathbb{N}$ to denote the
number of factors in the object $x$, when evaluated on the required
contextual level. For example, in the case of real numbers, we have $F\left[
4\right] =1$ but $F\left[ 2\times 2\right] =2$.

The notation $\mathcal{H}_{[12\ldots N]}$ denotes a quantum register of rank
$N$, consisting of the tensor product
\begin{equation}
\mathcal{H}_{\left[ 12\ldots N\right] }\equiv \mathcal{H}_{1}\otimes
\mathcal{H}_{2}\otimes \ldots \otimes \mathcal{H}_{N}  \label{111}
\end{equation}
of a finite number $N\geqslant 1 $ of factor Hilbert spaces $\mathcal{H}_{i}$%
, $1\leqslant i\leqslant N$, each known as a (quantum) subregister. The
dimension $d_{i}$ of the $i^{th}$ subregister $\mathcal{H}_{i}$ will
generally be assumed to be finite. When this dimension is two, such a
subregister is called a quantum bit, or \emph{qubit}. Note that $F\left[%
\mathcal{H}_{\left[ 12\ldots N\right] }\right] =1$ but $F\left[ \mathcal{H}%
_{1}\otimes \mathcal{H}_{2}\otimes\ldots \otimes \mathcal{H}_{N}\right] =N$.
This is an example where an equality is \emph{contextual}, that is, does not
hold under all circumstances.

The left-right ordering of the tensor product in $\left( \ref{111}\right) $
is not significant in our approach. Left-right ordering turns out to be an
inadequate way of labelling products in the case of three or more
subregisters, because entanglements can occur between elements of any of the
subregisters. Instead, we shall use subscript labels as in (\ref{111}) to
identify specific subregisters, which are therefore regarded as having their
own physical identities. For example,
\begin{equation}
\mathcal{H}_{\left[ 12\right] }\equiv \mathcal{H}_{1}\otimes \mathcal{H}%
_{2}= \mathcal{H}_{2}\otimes \mathcal{H}_{1}\equiv \mathcal{H}_{\left[ 21%
\right] }.
\end{equation}

This invariance to left-right re-ordering applies to states of subregisters
as well as the subregisters themselves. For example, if $\psi _{1}\in H_{1}$
and $\phi _{2}\in H_{2}$, then
\begin{equation}
\phi _{2}\otimes \psi _{1}=\psi _{1}\otimes \phi _{2}.
\end{equation}

\

\begin{center}
\textbf{A. Splits}
\end{center}

A quantum register consisting of two or more subregisters can be \emph{split}
in a number of ways. A split is just an arrangement of the subregisters into
a convenient number of nonintersecting groupings of tensor products. For
example, a rank-$3$ quantum register can be split in five different ways:
\begin{equation}
\mathcal{H}_{[123]}\equiv \mathcal{H}_{\left[ 12\right] }\otimes \mathcal{H}%
_{3}=\mathcal{H}_{\left[ 13\right] }\otimes \mathcal{H}_{2}=\mathcal{H}_{%
\left[ 23\right] }\otimes \mathcal{H}_{1}=\mathcal{H}_{1}\otimes \mathcal{H}%
_{2}\otimes \mathcal{H}_{3}.
\end{equation}
The equality here refers to the fact that each of these splits is the same
as a vector space, but they are not equivalent in terms of split structure.
For example, $F\left[ \mathcal{H}_{[123]}\right] =1$ but $F\left[ \mathcal{H}%
_{\left[ 12\right] }\otimes \mathcal{H}_{3}\right] =2$ and $F\left[ \mathcal{%
H}_{1}\otimes \mathcal{H}_{2}\otimes \mathcal{H}_{3}\right] =3$. The
significance of such splits is that states which are factorizable relative
to one split need not be factorizable relative to another \cite
{JAROSZKIEWICZ-03A}. This underlines the fact that entanglement and
factorization are context dependent, that is, depend on the physical
interpretation of the subregisters concerned.

Although they are crucial to the development of quantum causal set theory
\cite{JAROSZKIEWICZ-03B}, splits by themselves do not go far enough to
describe physics, however, and we need to develop the notion of a \emph{%
partition}, which is based on the concepts of \emph{separations} and \emph{%
entanglement}, discussed next.

\

\begin{center}
\textbf{B. Separations}
\end{center}

We define the separations first because entanglements can only be defined in
terms of them. Given a rank-$N$ quantum register $\mathcal{H}_{[12\ldots N]}$%
, each of its component subregisters $\mathcal{H}_{i}$, $1\leqslant
i\leqslant N$\ will be called a \emph{rank-1} subregister. Rank-1
subregisters will be assumed to be elementary, in that the concepts of
entanglements and separations $($defined below$)$\ do not apply to them.
Qubits are examples of such elementary subregisters. An arbitrary element in
a subregister $\mathcal{H}_{i}$ will be denoted by $\psi _{i}$, except when
the index is the letter $n$, in which case it refers to exotime.

Assuming $N>1$, for any choice of two rank-1 subregisters $\mathcal{H}_{i},%
\mathcal{H}_{j}$ in $\left( \ref{111}\right) $ such that $1\leqslant i <
j\leqslant N$, we define the \emph{rank-2 subregister} $\mathcal{H}_{[ij]} =
\mathcal{H}_{[ji]}$ of $\mathcal{H}_{\left[ 12\ldots N\right] }$ as the
tensor product
\begin{equation}
\mathcal{H}_{[ij]}\equiv \mathcal{H}_{i}\otimes \mathcal{H}_{j},
\end{equation}
there being a total of $\frac{_{1}}{^{2}}N\left( N-1\right) $ distinct
rank-2 subregisters. Rank-2 subregisters can contain both entangled and
separable states and are vector spaces in their own right. Consistent with
our notation, an element in $\mathcal{H}_{[ij]}$\ will be denoted by $\psi _{%
\left[ ij\right] }$.

For a given rank-2 subregister $\mathcal{H}_{\left[ ij\right] }$, we define
the \emph{rank-2 separation} $\mathcal{H}_{ij}$ to be the proper subset of $%
\mathcal{H}_{\left[ ij\right] }$ consisting of all separable elements in it,
i.e.,
\begin{equation}
\mathcal{H}_{ij}\equiv \left\{ \psi _{i}\otimes \phi _{j}:\left( \psi
_{i}\in \mathcal{H}_{i}\right) \;\&\;\left( \phi _{j}\in \mathcal{H}%
_{j}\right) \right\} .
\end{equation}
In this notation, the subscripts are not basis set indices. By definition we
include in $\mathcal{H}_{ij}$ the zero vector $0_{[ij]}$ of $\mathcal{H}%
_{[ij]}$.

We shall use lower indices without square brackets to denote separations,
reserving the use of lower indices within square brackets to represent
tensor products of subregisters. The concept of rank-2 separation
generalizes readily to higher rank separations \cite{JAROSZKIEWICZ-03B}. For
example, the rank-3 separation $\mathcal{H}_{ijk}$ is the subset of $%
\mathcal{H}_{\left[ ijk\right] }$ defined by
\begin{equation}
\mathcal{H}_{ijk}\equiv \left\{ \phi _{i}\otimes \psi _{j}\otimes \mathcal{%
\eta }_{k}:\left( \phi _{i}\in \mathcal{H}_{i}\right) \&\left( \psi _{j}\in
\mathcal{H}_{j}\right) \&\left( \eta _{k}\in \mathcal{H}_{k}\right) \right\}
.
\end{equation}

\

\begin{center}
\textbf{C. Entanglements}
\end{center}

Entanglements may be constructed once the separations have been defined.
Starting with the lowest rank possible, we define the \emph{rank-2
entanglement} $\mathcal{H}^{ij}$\ to be the complement of $\mathcal{H}_{ij}$
in $\mathcal{H}_{[ij]}$, i.e.,
\begin{equation}
\mathcal{H}^{ij}\equiv \mathcal{H}_{\left[ ij\right] }-\mathcal{H}%
_{ij}=\left( \mathcal{H}_{\left[ ij\right] }\mathcal{\cap H}_{ij}\right)
^{c}.
\end{equation}
Hence $\mathcal{H}_{\left[ ij\right] }=\mathcal{H}_{ij}\cup \mathcal{H}^{ij}$%
. $\mathcal{H}_{ij}$ and $\mathcal{H}^{ij}$ are disjoint and $\mathcal{H}%
^{ij}$\ does not contain the zero vector. An important aspect of this
decomposition is that neither $\mathcal{H}_{ij}$ nor $\mathcal{H}^{ij}$ is a
vector space.

The generalization of the entanglements to higher rank subregisters is
straightforward but requires the concept of \emph{separation product}. If $%
\mathcal{A}_{i}$ and $\mathcal{B}_{j}$ are arbitrary, non-empty subsets of $%
\mathcal{H}_{i}$ and $\mathcal{H}_{j}$ respectively, where $i\neq j$, then
we define the separation product $\mathcal{A}_{i}\bullet \mathcal{B}_{j}$ to
be the subset of $\mathcal{H}_{[ij]}$ given by
\begin{equation}
\mathcal{A}_{i}\bullet \mathcal{B}_{j}\equiv \left\{ \psi _{i}\otimes \phi
_{j}:\psi \in \mathcal{A}_{i},\;\phi \in \mathcal{B}_{j}\right\} .
\end{equation}
This generalizes immediately to any sort of product. For example, $\mathcal{H%
}_{ij}=\mathcal{H}_{i}\bullet \mathcal{H}_{j}.$ Separation products are
associative, commutative and cumulative, i.e.,
\begin{eqnarray}
\left( \mathcal{H}_{i}\bullet \mathcal{H}_{j}\right) \bullet \mathcal{H}_{k}
&=&\mathcal{H}_{i}\bullet \left( \mathcal{H}_{j}\bullet \mathcal{H}%
_{k}\right) \equiv \mathcal{H}_{ijk}  \notag \\
\mathcal{H}_{ij}\bullet \mathcal{H}_{k} &=&\mathcal{H}_{ijk},
\end{eqnarray}
and so on. Separation products can also be defined for the entanglements.
For example,
\begin{eqnarray}
\mathcal{H}^{ij}\bullet \mathcal{H}_{k} &=&\left\{ \phi ^{ij}\otimes \psi
_{k}:\phi ^{ij}\in \mathcal{H}^{ij},\;\psi _{k}\in \mathcal{H}_{k}\right\} ,
\notag \\
\mathcal{H}^{ij}\bullet \mathcal{H}^{rs} &=&\left\{ \phi ^{ij}\otimes \psi
^{rs}:\phi ^{ij}\in \mathcal{H}^{ij},\psi ^{rs}\in \mathcal{H}^{rs}\right\} .
\end{eqnarray}
A further notational simplification is to use a single $\mathcal{H}$ symbol,
using the vertical position of indices to indicate separations and
entanglements and incorporating the separation product symbol $\bullet $ with
indices directly. For example,
\begin{equation}
\mathcal{H}^{15}\bullet \mathcal{H}^{97}\bullet \mathcal{H}_{28}\bullet
\mathcal{H}_{4}\bullet \mathcal{H}_{36}\equiv \mathcal{H}_{28\bullet
4\bullet 36}^{15\bullet 97}=\mathcal{H}_{23468}^{15\bullet 97}.
\end{equation}
Associativity of the separation product applies to both separations and
entanglements.

Rank-3 and higher entanglements such as $\mathcal{H}^{ijkl}$ are defined in
terms of complements. For example, we define
\begin{equation}
\mathcal{H}^{abc}\equiv \mathcal{H}_{[abc]}-\{\mathcal{H}_{abc}\cup \mathcal{%
H}_{a}^{bc}\cup \mathcal{H}_{b}^{ac}\cup \mathcal{H}_{c}^{ab}\}.
\end{equation}

\

\begin{center}
\textbf{D. Partitions}
\end{center}

Every quantum register $\mathcal{H}$ can be represented uniquely as a union
of disjoint elements, such as separations, entanglements, and separation
products of these two types (such as $\mathcal{H}_{a}^{bcd}\equiv \mathcal{H}%
_{a}\bullet \mathcal{H}^{bcd}$). These elements will be called \emph{%
partitions} and together they form the \emph{natural lattice} $\frak{L}%
\left( \mathcal{H}\right) $ of $\mathcal{H}$. The number of elements in each
natural lattice is given by the Bell numbers \cite{JAROSZKIEWICZ-03B}. For
example, a rank-$3$ quantum register has five partitions:
\begin{equation}
\mathcal{H}_{[abc]}=\mathcal{H}_{abc}\cup \mathcal{H}_{a}^{bc}\cup \mathcal{H%
}_{b}^{ac}\cup \mathcal{H}_{c}^{ab}\cup \mathcal{H}^{abc}.
\end{equation}

Each partition itself may be the separation product of a number of \emph{%
blocks}, each block being an individual separation or entanglement. For
example, the partition $\mathcal{H}_{cd}^{ab\bullet efg}$ has four blocks:
two separations, $\mathcal{H}_{c}$, $\mathcal{H}_{d}$ and two entanglements,
$\mathcal{H}^{ab}$, $\mathcal{H}^{efg}$.

\

We may also use the above index notation to label the various elements of
the entanglements and separations. For example, $\psi _{abc}^{def\;\bullet
gh}$\ is interpreted to be some element in the partition $\mathcal{H}%
_{abc}^{def\;\bullet gh}$ and so on. With this notation we may write for
example
\begin{equation}
\psi _{abc}^{def\;\bullet gh}=\psi _{a}\otimes \psi _{b}\otimes \psi
_{c}\otimes \psi ^{def}\otimes \psi ^{gh},
\end{equation}
where $\psi _{a}\in \mathcal{H}_{a},\;\psi _{b}\in \mathcal{H}_{b}$, $\psi
_{c}\in \mathcal{H}_{c}$, $\psi ^{def}$\ $\in \mathcal{H}^{def}$\ and $\psi
^{gh}\in \mathcal{H}^{gh}$. Each factor such as $\psi _{a},\psi ^{def},etc.$
lies in a particular block in the partition to which $\psi
_{abc}^{def\;\bullet gh}$ belongs.

An important feature of the concepts of splits, separations, entanglements
and partitions is that they are all independent of basis, the only
requirement for their definition being that the enclosing Hilbert space is a
tensor product of identifiable subregisters.

\

\begin{center}
{\large {$\mathbf{V.}$ \textbf{OPERATORS}} }
\end{center}

In the stages paradigm, operators representing tests are assumed to be
Hermitian because this guarantees that their eigenvalues are real and that
nondegenerate eigenstates are orthogonal. In general, for a finite
dimensional Hilbert space $\mathcal{H}$ of dimension $d$ we can find $d^{2}$
linearly independent Hermitian operators out of which we can build all the
other Hermitian operators on $\mathcal{H}$ \cite{PERES:93}. These
independent operators can then be used as a basis for the real vector space $%
\mathbb{H}\left( \mathcal{H}\right) $ of all Hermitian operators on $%
\mathcal{H}$. Furthermore, multiplication of elements in $\mathbb{H}\left(
\mathcal{H}\right) $ by other elements in $\mathbb{H}\left( \mathcal{H}%
\right) $ is well defined and closed, so that $\mathbb{H}\left( \mathcal{H}%
\right) $ is also an algebra \cite{HOWSON:72}\ over the real number field.

Given a quantum register consisting of $N$ subregisters, we can go further
and define \emph{skeleton sets of operators} \cite{JAROSZKIEWICZ-03B}. A
skeleton set is a basis for $\mathbb{H}\left( \mathcal{H}_{\left[ 1\ldots N%
\right] }\right) $ such that every element of the set is factorizable into a
product of $N$ factors, the $n^{th}$ factor being associated with the $%
n^{th} $ subregister. To construct such a skeleton set, we first construct
an operator basis for each algebra $\mathbb{H}\left( \mathcal{H}_{i}\right)
,\;1\leqslant i\leqslant N$, and then take tensor products of elements of
bases from all the different subregisters. For example, for a two qubit
register $\mathcal{H}_{\left[ 12\right] }$, a skeleton set for $\mathbb{H}%
\left( \mathcal{H}_{\left[ 12\right] }\right) $ is given by the sixteen
elements $\left\{ \hat{\sigma}_{1}^{\mu }\otimes \hat{\sigma}_{2}^{\nu
}:0\leqslant \mu ,\nu \leqslant 3\right\} $ where $\hat{\sigma}_{i}^{0}$ is
the identity operator in $\mathcal{H}_{i}$ and $\left\{ \hat{\sigma}_{i}^{1},%
\hat{\sigma}_{i}^{2}, \hat{\sigma}_{i}^{3}\right\} $ are equivalent to Pauli
spin matrices.

Skeleton sets of operators permit us to define separations and entanglements
for $\mathbb{H}\left( \mathcal{H}_{\left[ 1\ldots N\right] }\right) $ in
much the same way as for the quantum register itself and we may use the same
index notation to represent separations and entanglements for operators as
for the states. For example, $\hat{A}_{1}^{23}$ will be understood to be an
operator of the form $\hat{A}_{1}\otimes \hat{B}^{12}$, where $\hat{A}_{1}$
is an element of $\mathbb{H}_{1}\equiv \mathbb{H}\left( \mathcal{H}%
_{1}\right) $ and $\hat{B}^{12}$ is an element of $\mathbb{H}^{12}$, which
is the set of all entangled elements of $\mathbb{H}\left( \mathcal{H}%
_{[12]}\right) ,$ i.e., those not of the form $\hat{C}_{2}\otimes \hat{D}%
_{3} $ where $\hat{C}_{2}\in \mathbb{H}_{2}$ and $\hat{D}\in \mathbb{H}_{3}$.

At first sight it may seem incorrect to apply the concept of entanglement to
Hermitian operators, given that in standard quantum mechanics they usually
represent physical observables corresponding to real physical laboratory
equipment. Such equipment appears to us to be rather classical and quite
correctly we would not normally think of applying the superposition
principle to it \emph{per se}. In the context we are discussing here,
however, superposition refers to vector addition in the abstract space $%
\mathbb{H}\left( \mathcal{H}\right) $ of operators representing physical
systems, and this will not in general translate directly to anything like
the ``addition'' of physical pieces of equipment together. If $\hat{\Sigma}%
_{1}$ and $\hat{\Sigma}_{2}$ are legitimate tests representing real physical
experiments $E_{1}$ and $E_{2}$ respectively, then $\hat{\Sigma}_{1}+\hat{%
\Sigma}_{2}$ \emph{might} also represent some other real physical experiment
$E_{3}$, but $E_{3}$ need not have anything to do with either $E_{1}$ or $%
E_{2}$ separately.

The study of the separation and entanglement properties of operators
representing tests has been relatively neglected in quantum theory at the
expense of the study of the separation and entanglement properties of
states, but it is clearly an important part of quantum dynamics
nevertheless. For instance, in the EPR thought experiment discussion of
entanglement \cite{EPR}, the spatially separated tests used to observe the
components of an entangled state are implicitly assumed to be represented by
factorizable operators (otherwise the spatially separated observers could
not make independent choices of what to measure), whilst Theorem 3
(discussed below) says that the test which had created the initial entangled
state had to be entangled.

Not only are the separation and entanglement properties of operators
important in their own right, but their relationship with the separation and
entanglement properties of states is an important and subtle one, leading to
a complex pattern of causal relationships which generates quantum causal set
structure \cite{JAROSZKIEWICZ-03B}. Fortunately, there are some simple yet
powerful theorems controlling the sort of outcomes we should expect from
various quantum tests. To understand these results, we should keep in mind
that the important structures in our paradigm are not the operators as such
but their associated basis sets. This was something recognized by Everett
\cite{EVERETT-57}.

\

\begin{center}
{\large {$\mathbf{VI.}$ \textbf{EIGENVALUES AND PREFERRED BASES}}}
\end{center}

In the following we shall assume all Hilbert spaces are finite dimensional
and make frequent references to the following terms:

$i)$ a \emph{degenerate} operator is a Hermitian operator with at least two
linearly independent eigenstates having identical eigenvalues of that
operator;

$ii)$ a \emph{weak} operator is a Hermitian operator which is either
degenerate or at least one of its eigenvalues is zero;

$iii)$ a \emph{strong} operator is a Hermitian operator which is not weak;
that is, none of its eigenvalues are zero and all its eigenvalues are
distinct.

The subset of $\mathbb{H}\left( \mathcal{H}\right) $\ consisting of all weak
elements of $\mathbb{H}\left( \mathcal{H}\right) $\ will be denoted by $%
\mathbb{W}\left( \mathcal{H}\right) $ whilst the subset of $\mathbb{H}\left(
\mathcal{H}\right) $ consisting of all strong elements of $\mathbb{H}\left(
\mathcal{H}\right) $ will be denoted by $\mathbb{S}\left( \mathcal{H}\right)$%
. Then clearly
\begin{equation}
\mathbb{S}\left( \mathcal{H}\right) \cup \mathbb{W}\left( \mathcal{H}\right)
=\mathbb{H}\left( \mathcal{H}\right) ,\;\;\;\mathbb{S}\left( \mathcal{H}%
\right) \cap \mathbb{W}\left( \mathcal{H}\right) =\emptyset .
\end{equation}
Neither $\mathbb{S}\left( \mathcal{H}\right) $ nor $\mathbb{W}\left(\mathcal{%
H}\right) $ are vector spaces.

\

We shall now discuss some important theorems involving strong and weak
operators which have significant implications for the physics associated
with quantum registers. Proofs of most of these are elementary and are
discussed in \cite{JAROSZKIEWICZ-03B}.

\

\begin{center}
\textbf{Theorem 1}
\end{center}

\begin{enumerate}
\item[\ ]  The normalized eigenstates of any strong operator $\hat{A}\in $\ $%
\mathbb{S}\left( \mathcal{H}\right) $ form a unique, orthonormal basis set
for $\mathcal{H}$, referred to as the \emph{preferred basis of} $\mathcal{H}$
\emph{relative to} $\hat{A}$, denoted by $\mathsf{B}_{A}\equiv \mathsf{B}%
_{A}\left( \mathcal{H}\right) $.
\end{enumerate}

We can use this theorem to relate the concepts of preferred basis sets and
strong operators as follows. First we note that associated with our
universal quantum register $\mathcal{H}$ is the set $\frak{B}\left( \mathcal{%
H}\right) $ of distinct orthonormal basis sets for $\mathcal{H}$. Second,
there is a many-to-one mapping from $\mathbb{S}\left( \mathcal{H}\right) $
onto $\frak{B}\left( \mathcal{H}\right) ,$ defined by virtue of Theorem 1.
Third, this mapping defines an equivalence relationship for elements in $%
\mathbb{S}\left( \mathcal{H}\right) $ which we call \emph{basis equivalence}%
: $\hat{A}$, $\hat{B}$ $\in \mathbb{S}\left( \mathcal{H}\right) $ are basis
equivalent if and only if
\begin{equation}
\mathsf{B}_{A}=\mathsf{B}_{B}.
\end{equation}
Basis equivalence is symmetric, transitive and reflexive and therefore
defines an equivalence relationship which divides $\mathbb{S}\left( \mathcal{%
H}\right) $ into disjoint equivalence classes. All the elements of a given
equivalence class have the same preferred basis.

\

\ Suppose now we have a rank-2 quantum register $\mathcal{H}_{\left[ 12%
\right] }\equiv \mathcal{H}_{1}\otimes \mathcal{H}_{2}$ and suppose $\hat{O}%
_{1}$\ $\in \mathbb{H}\left( \mathcal{H}_{1}\right) $\ and $\hat{O}_{2}$\ $%
\in \mathbb{H}\left( \mathcal{H}_{2}\right) $. Then the tensor product
operator $\hat{O}_{12}\equiv \hat{O}_{1}\otimes \hat{O}_{2}$\ is a separable
element of $\mathbb{H}\left( \mathcal{H}_{[12]}\right) $ and the following
theorem holds:

\

\begin{center}
\textbf{Theorem 2}
\end{center}

\begin{enumerate}
\item[i)]  If $\hat{O}_{1}$\ or $\hat{O}_{2}$\ is weak then $\hat{O}%
_{12}\equiv $\ $\hat{O}_{1}\otimes \hat{O}_{2}$ is necessarily weak;

\item[ii)]  equivalently, a tensor product operator is strong only if each
of its factors is strong;

\item[iii)]  if both $\hat{O}_{1}$\ and $\hat{O}_{2}$\ are strong, $\hat{O}%
_{12}$\ need not be strong.
\end{enumerate}

Theorem 2 leads to the following theorem which has important implications
for the outcomes of certain kinds of physics experiments:

\

\begin{center}
\textbf{Theorem 3 (the fundamental theorem)}
\end{center}

\begin{enumerate}
\item[i)]  All the eigenstates of a separable strong operator are separable;

\item[ii)]  equivalently, entangled states can be the outcomes of entangled
operators only.
\end{enumerate}

\

These results generalize to higher rank quantum registers. The importance of
Theorem 3 is that it forces the factorization properties of the tests to
drive the factorization properties of the states. It is this mechanism which
underpins our analysis of quantum causal set theory \cite{JAROSZKIEWICZ-03B}%
. To see how this works, consider an initial state of the universe $\Psi _{n}
$ which has precisely $a$ factors, i.e.,
\begin{equation}
\Psi _{n}=\psi _{\left( 1\right) }\otimes \psi _{\left( 2\right) }\otimes
\ldots \otimes \psi _{\left( a\right) },
\end{equation}
where $\psi _{\left( k\right) }$ is a completely entangled element in some
rank-$r_{k}$ factor of a particular split $\mathcal{S}_{1}$ of the universal
register $\mathcal{H}_{\left[ 1\ldots N\right] }$. By relabelling the
subregisters appropriately, we may always write this split in the form
\begin{equation}
\mathcal{S}_{1}=\mathcal{H}_{\left[ 1\ldots r_{1}\right] }\otimes \mathcal{H}%
_{\left[ \left( r_{1}+1\right) \ldots \left( r_{1}+r_{2}\right) \right]
}\otimes \ldots \otimes \mathcal{H}_{\left[ \left( N+1-r_{a}\right) \ldots N%
\right] }.
\end{equation}
Now suppose that the next test of the universe $\hat{\Sigma}_{n+1}$\
factorizes into precisely $b$ factors, i.e.,
\begin{equation}
\hat{\Sigma}_{n+1}=\hat{\sigma}_{\left( 1\right) }\otimes \hat{\sigma}%
_{\left( 2\right) }\ldots \hat{\sigma}_{\left( b\right) },
\end{equation}
where each of the factors $\hat{\sigma}_{\left( l\right) }$ is a completely
entangled strong operator acting over some rank-$s_{l}$ factor of another
split $\mathcal{S}_{2}\left( \mathcal{H}_{\left[ 1\ldots N\right] }\right) $
of the universal register. In general, it will not be possible to easily
relate the two splits $\mathcal{S}_{1}$, $\mathcal{S}_{2}$ subregister by
subregister, because for any two randomly chosen splits of the same
register, the likelihood is that none of their factors co-incide. It is this
which creates the rich structure of quantum causal set theory \cite
{JAROSZKIEWICZ-03B}.

Now according to Theorem 3, the outcome $\Psi _{n+1}$ of test $\Sigma _{n+1}$
has to factorize into $c$ factors, where $c\geqslant b$. These factors can
be arranged into precisely $b$ groups of factors, i.e.
\begin{equation}
\Psi _{n+1}=\phi _{\left( 1\right) }\otimes \phi _{\left( 2\right) }\otimes
\ldots \otimes \phi _{\left( b\right) },
\end{equation}
where each group of factors $\phi _{\left( t\right) }$ defines a factor
state of $\Psi _{n+1}$ which is an eigenstate of $\hat{\sigma}_{\left(
t\right) }$ and therefore lies in the corresponding factor of the split $%
\mathcal{S}_{2}$. It is possible for each of these $b$ groups of factors to
consist of one or more factors, because this is not ruled out by the theorem
(entangled strong operators can have separable outcomes, but not the other
way around).

The relevance to quantum cosmology of these observations is that, \emph{if}
successive tests of the universe contain progressively more factors, then
successive states of the universe must factorize at least to the same
extent. If now we identify factorization of states with \emph{classicality (}%
the appearance of distinct physical identity), then an explanation for the
expansion of the universe could be that it is driven by the increasing
factorization properties of successive tests of the universe.

\

\begin{center}
\textbf{A. Quantum zipping}
\end{center}

A fundamental feature of quantum register dynamics which is central to all
of this discussion is that inner products between successive states have to
follow the rule that, whatever the details of these states and regardless of
which split each is in, individual subregister component states always have
to ``zip'' together in inner products. For example, for a rank-2 quantum
register, if $\mathsf{B}\equiv \left\{ |i\rangle _{1}\otimes |j\rangle
_{2}\right\} $ and $\mathsf{B}^{\prime }\equiv \left\{ |a\rangle _{1}\otimes
|b\rangle _{2}\right\} $ are two factorizable basis sets for the register,
the inner product $\left( \Phi ,\Psi \right) $ of states $\Psi \equiv
\sum_{i,j}\psi _{ij}|i\rangle _{1}\otimes |j\rangle _{2}$ and $\Phi \equiv
\sum_{a,b}\phi _{ab}|a\rangle _{1}\otimes |b\rangle _{2}$ is of the form
\begin{equation}
\left( \Phi ,\Psi \right) =\sum_{i,a}\sum_{j,b}\phi _{ab}^{\ast }\psi
_{ij}\langle a|i\rangle _{1}\langle b|j\rangle _{2}.
\end{equation}
Quantum zipping is discussed in more detail in $\S XI$.

\

\begin{center}
{\large {$\mathbf{VII.}$ \textbf{NATURAL BASES}} }
\end{center}

Given two finite dimensional Hilbert spaces $\mathcal{H}_{1}$, $\mathcal{H}%
_{2}$\ with dimensions $d_{1},d_{2}$\ respectively, then their tensor
product $\mathcal{H}_{\left[ 12\right] }\equiv \mathcal{H}_{1}\otimes
\mathcal{H}_{2}$\ is also a finite dimensional Hilbert space with dimension $%
d_{\left[ 12\right] }=d_{1}d_{2}.$\ However, $\mathcal{H}_{\left[ 12\right]
} $\ is more than just a Hilbert space with dimension $d_{\left[ 12\right] }$%
. Because it is a tensor product, it is possible to discuss separations and
entanglements as explained above, which cannot be done with a Hilbert space
with the same dimension but not known to be a tensor product. This is an
elementary fact which is fundamental for the development of our view of the
quantum universe. It arises precisely because a tensor product has component
spaces (in this example $\mathcal{H}_{1}$\ and $\mathcal{H}_{2}$) which have
their own identities. In other words, a quantum register exists as such
simply because it has identifiably distinct subregisters.

This identifiability of subregisters is unrelated to the concepts of \emph{%
distinguishability }and \emph{indistinguishability} of particles in quantum
mechanics. Jordan and Wigner showed a long time ago that quantum registers
based on the principles we employ here can be used to construct quantum
fields with bosonic or fermionic symmetry \cite
{JORDAN+WIGNER-28,BJORKEN+DRELL:65B}.

Another important aspect of this discussion is that the concepts of splits,
separations, entanglements and partitions are defined without reference to
any basis sets, either for the subregisters or for quantum register itself.
For example, given that $\Psi \in \mathcal{H}_{\left[ 12\right] }$\ is
separable, then we can be sure that there exist elements $\psi _{1}\in
\mathcal{H}_{1}$\ and $\psi _{2}\in \mathcal{H}_{2}$\ such that $\Psi =\psi
_{1}\otimes \psi _{2}$, this statement being independent of any choice of
basis for any of the spaces concerned.

This point is more subtle than its first appears, because it is possible to
find basis sets for the tensor product space $\mathcal{H}_{[12]}$\ such that
all their elements are entangled. Somewhat surprisingly, such a basis set
can then be used to describe separable states. A discussion of this and
related concepts is given in \cite{JAROSZKIEWICZ-03A}. There we discussed
the possibility of finding basis sets for tensor product spaces which have
mixed factorization properties. For example, in the case of $\mathcal{H}_{%
\left[ 12\right] }$, a basis of type $\left( p,q\right) $\ has $p$\ elements
which are entangled (i.e., are elements of $\mathcal{H}^{12})$\ and $%
q=d_{1}d_{2}-p$\ which are separable (i.e., are elements of $\mathcal{H}%
_{12}).$

In \cite{JAROSZKIEWICZ-03A} we defined a \emph{completely entangled basis}
for $\mathcal{H}_{\left[ 12\right] }$\ to be one of type $\left(
d_{1}d_{2},0\right) $, whereas a \emph{completely separable basis} is of
type $\left( 0,d_{1}d_{2}\right) .$\ A completely separable basis will also
be referred to as a \emph{natural basis}, because such a basis displays the
full separability properties underlying a tensor product space, whereas the
other types either partially or completely hide these properties. In
particular, the use of a completely entangled basis is equivalent (on a
formal level) to replacing the tensor product space $\mathcal{H}_{[12]}$\
with a featureless Hilbert space which happens to have dimension $%
d_{1}d_{2}. $\ As we have stated, however, matters are more subtle than they
appear. The reason we can use a completely entangled basis to discuss
separable states is that we still need to have an underlying knowledge of
the component spaces of $\mathcal{H}_{\left[ 12\right] }$\ in order to
define entanglement in the first place and it is this knowledge allows us to
say what we mean by ``separable'' state.

We shall use the symbol $\frak{B}\left( p,q\right) $\ to denote the set of
all type $\left( p,q\right) $\ bases for a rank-2 quantum register.

\

\begin{center}
{\large {$\mathbf{VIII.}$ \textbf{ACTIVE VERSUS PASSIVE TRANSFORMATIONS}} }
\end{center}

Mathematicians are generally concerned with functions, transformations and
mappings in a precise way and it is somewhat surprising therefore that the
mathematician's concept of a transformation needs to be qualified when it
comes to physics. In physics, it is most important to distinguish between
the concepts of \emph{passive} and \emph{active} transformations.

Passive transformations are purely formal changes in the \emph{descriptions}
of mathematical structures representing physical systems, one of the main
characteristics of these changes being that they are applied to all elements
of a set. By definition, passive transformations have no physical or
observable consequences.

An example of a passive transformation is a change from one spatial
co-ordinate frame of reference to another. Normally, such a change is
regarded by physicists has having no intrinsic effect on the measurable
physical relationships between material objects embedded in the space. This
idea becomes of the greatest importance in theories such as general
relativity, where a main objective is to identify those aspects of the
theory which are generally covariant\emph{, }i.e., are\emph{\ }invariant to
arbitrary co-ordinate transformations.

Another example of a passive change is a unitary transformation which acts
on all the elements in a Hilbert space. Such a transformation leaves all
inner products invariant and on account of this is regarded as physically
undetectable.

In contrast to the mathematician's passive transformation, a real physics
experiment always involves a physical change in some parts of the universe
and not in others. Locality is one of the characteristics of an active
transformation. An active transformation never acts on the universe as a
whole, because this would also include the physicists performing the
experiment. Clearly, the essential difference between active and passive
transformations involves the issue of exophysics versus endophysics.

Active and passive transformations are readily confused when the nonglobal
nature of an active transformation is overlooked. In such cases, sign
changes may be the only way of seeing the difference, as in the case of
spatial rotations. An analogous situation arises in quantum cosmology when
the role of time is considered. In the many-universe scenario, the state of
the universe $\Psi $ is assumed to satisfy the Schr\"{o}dinger equation
\begin{equation}
i\hbar \partial _{t}\Psi =\hat{H}\Psi ,  \label{444}
\end{equation}
which gives rise to the unitary evolution conventionally regarded as an
essential ingredient of quantum cosmology. Now this equation is given in the
Schr\"{o}dinger picture, in which operators are usually time independent. It
is well known however that alternative pictures can be used. In particular,
the Heisenberg picture may be used to freeze the quantum state, locating the
intrinsic time dependence within the operators representing the observables.
The interpretation of this is that the semiclassical observers assumed to be
present in the standard view of quantum mechanics carry an intrinsic (i.e.
physical) time dependence which is not removed by the passive unitary
transformation from Schr\"{o}dinger picture to Heisenberg picture. When
these observers decide to perform a measurement at a given moment $T$ of
their time, the corresponding Heisenberg picture operator representing the
said measurement carries a memory of the observer's time $T.$

From this we arrive at the standard equality of expectation values between
each of these two pictures:
\begin{equation}
\langle \Psi ,T|\hat{A}_{s}|\Psi ,T\rangle _{s}=\langle \Psi |\hat{A}%
_{H}\left( T\right) |\Psi \rangle _{H},
\end{equation}
where $|\Psi ,T\rangle _{S}=\hat{U}\left( T\right) |\Psi \rangle _{H}$, $%
\hat{A}_{H}\left( T\right) \equiv U^{+}\left( T\right) \hat{A}_{s}\hat{U}%
\left( T\right) $ and $\hat{U}\left( T\right) $ is the unitary temporal
evolution operator associated with $\left( \ref{444}\right)$. The big
problem for quantum cosmology is that it is generally assumed that there are
no external observers in the first place. Therefore, there is no form of
external memory of when any measurement is taken, which is why time itself
seems to have been transformed away by the change from the Schr\"{o}dinger
picture to the Heisenberg picture in quantum cosmology. A similar line of
argument, associated with general covariance, is behind the lack of a global
time in the Wheeler-de Witt equation
\begin{equation}
\hat{H}|\Psi \rangle =0.
\end{equation}

We are at risk of a similar problem arising in the stages paradigm, because
we too do not have any external observers. To avoid it, we must ensure that
our endophysical definition of information exchange is one which cannot be
undermined by any passive transformation of any sort. Such a definition can
be found based on the concept of partition change. The next two sections set
the scene for our statement of this definition.

\

\begin{center}
{\large {$\mathbf{IX.}$ \textbf{LOCAL TRANSFORMATIONS}} }
\end{center}

Given a Hilbert space $\mathcal{H}$, we define $\mathbb{U}\left( \mathcal{H}%
\right) $\ to be the set of all unitary transformations on $\mathcal{H}$.
Similarly, if $\mathcal{H}_{1}$\ and $\mathcal{H}_{2}$\ are two Hilbert
spaces, then we denote the set of all unitary transformations on their
tensor product $\mathcal{H}_{\left[ 12\right] }$\ by the symbol $\mathbb{U}_{%
\left[ 12\right] }\equiv \mathbb{U}\left( \mathcal{H}_{\left[ 12\right]
}\right) $.

Before we discuss local unitary operators in more detail, we need to clarify
one small point; $\mathbb{U}_{1}\equiv \mathbb{U}\left( \mathcal{H}%
_{1}\right) $ and $\mathbb{U}_{2}\equiv \mathbb{U}\left( \mathcal{H}%
_{1}\right) $ act on different spaces and are therefore unrelated.
Technically, neither is a subset of $\mathbb{U}_{[12]}$, for example.
However, whenever it suits our purposes, we shall assume without further
comment that when we write $\mathbb{U}_{1}$, for example, we may also mean $%
\mathbb{U}_{1}\otimes \hat{I}_{2}$, where $\hat{I}_{2}$\ is the identity
operator on $\mathcal{H}_{2}$, and so on, depending on context.

An important subset of $\mathbb{U}_{\left[ 12\right] }$\ is the set $\mathbb{%
U}_{12}\equiv \mathbb{U}_{1}\bullet \mathbb{U}_{2}$ of \emph{local unitary
transformations}, all the elements of which are of the form
\begin{equation}
\hat{U}_{12}\equiv \hat{U}_{1}\otimes \hat{U}_{2},\;\;\;\hat{U}_{i}\in
\mathbb{U}_{i},\;\;\;\ i=1,2.
\end{equation}

The local unitary transformations form a nonabelian group under operator
product multiplication.

The significance of local unitary transformations is that they transform
natural bases into other natural bases, as can be readily proved. For
example, a natural basis $\mathsf{B}_{\left( 0,d_{1}d_{2}\right) }\in \frak{B%
}_{\left( 0,d_{1}d_{2}\right) }$ has elements of the form $\phi _{12}\equiv
a_{1}\otimes b_{2}\in \mathcal{H}_{12}$, where $a_{1}\in \mathcal{H}_{1}$
and $b_{2}\in \mathcal{H}_{2}$. Then a local unitary transformation $\hat{U}%
_{12}$ of each such element $\phi _{12}$\ gives
\begin{eqnarray}
\hat{U}_{12}\phi _{12} &=& ( \hat{U}_{1}\otimes \hat{U}_{2}) ( a_{1}\otimes
b_{2} )  \notag \\
&=& ( \hat{U}_{1}a_{1} ) \otimes ( \hat{U}_{2}b_{2})  \notag \\
&=&\phi _{12}^{\prime }\in \mathcal{H}_{12}.
\end{eqnarray}
This, and the fact that unitary transformations preserve inner products,
proves the assertion. Formally, we shall write
\begin{equation}
\mathbb{U}_{12}\frak{B}_{\left( 0,d_{1}d_{2}\right) }=\frak{B}_{\left(
0,d_{1}d_{2}\right) }.
\end{equation}
This generalizes in an obvious way to higher rank tensor product spaces.

This leads to the following theorem which is important in our concept of
information exchange:

\

\begin{center}
\textbf{Theorem 4}
\end{center}

\begin{enumerate}
\item[\ ]  Local unitary transformations are invariances of separations and
entanglements.
\end{enumerate}

\begin{center}
\textbf{Proof}
\end{center}

We prove this first for separations and then for entanglements:

\begin{enumerate}
\item[i)]  \qquad \textbf{Separations}: let $\hat{U}_{1\ldots n}\equiv \hat{U%
}_{1}\otimes \ldots \otimes \hat{U}_{n}\;$be any element of $\mathbb{U}%
_{1\ldots n}$\ and let $\phi _{1\ldots n}\equiv \phi _{1}\otimes \ldots
\otimes \phi _{n}$\ be any element of $\mathcal{H}_{1\ldots n}$. Then
\begin{eqnarray}
\hat{U}_{1\ldots n}\phi _{1\ldots n} &=&\left( \hat{U}_{1}\otimes \ldots
\otimes \hat{U}_{n}\right) \left( \phi _{1}\otimes \ldots \otimes \phi
_{n}\right)  \notag \\
&=&\left( \hat{U}_{1}\phi _{1}\right) \otimes \ldots \left( \hat{U}_{n}\phi
_{n}\right) ,\;\;\;\in \mathcal{H}_{1\ldots n}.
\end{eqnarray}
Hence
\begin{equation}
\mathbb{U}_{1\ldots n}\mathcal{H}_{1\ldots n}=\mathcal{H}_{1\ldots n}
\end{equation}
as required.

\item[ii)]  \qquad \textbf{Entanglements}: Let $\hat{U}_{1\ldots m}\equiv
\hat{U}_{1}\otimes \ldots \otimes \hat{U}_{m}\;$ be any element of $\mathbb{U%
}_{1\ldots m}$\ and let $\phi ^{1\ldots m}$\ be any element of the
entanglement $\mathcal{H}^{1\ldots m}$\ $(m>1)$. Now suppose that $\hat{U}%
_{1\ldots m}$\ takes $\phi ^{1\ldots m}$\ out of $\mathcal{H}^{1\ldots m}$
into some state $\phi ^{\prime }$ not in $\mathcal{H}^{1\ldots m}$. Because $%
\phi ^{\prime }$ is given as not in $\mathcal{H}^{1\ldots m},$ it is
necessarily in the complement $\mathcal{H}_{[1\ldots m]}-\mathcal{H}%
^{1\ldots m}$ of $\mathcal{H}^{1\ldots m}$ in the register $\mathcal{H}%
_{[1\ldots m]}.$ Then because $m>1$ and by definition and construction of
entanglements, the result $\phi ^{\prime }$ must necessarily be a separable
state, i.e., we may write
\begin{equation}
\hat{U}_{1\ldots m}\phi ^{1\ldots m}=\Phi \otimes \Psi \in \mathcal{H}%
_{[1\ldots m]}-\mathcal{H}^{1\ldots m},  \label{201}
\end{equation}
for some factor states $\Phi $, $\Psi $.

Now without loss of generality we may always relabel the subregisters so
that we may write
\begin{equation}
\Phi \in \mathcal{H}_{\left[ 1\ldots k\right] },\;\;\;\Psi \in \mathcal{H}_{%
\left[ (k+1)\ldots m\right] },
\end{equation}
for some $k$\ satisfying the condition $1\leqslant k<m$.

Next, because $\hat{U}_{1\ldots m}$\ is a unitary operator, it has an
inverse, $\hat{U}_{1\ldots m}^{-1}$, given by
\begin{equation}
\hat{U}_{1\ldots m}^{-1}=\hat{U}_{1}^{-1}\otimes \ldots \otimes \hat{U}%
_{m}^{-1}.
\end{equation}
This inverse is also a local unitary transformation. Applying this inverse
to equation $\left( \ref{201}\right) $\ gives
\begin{eqnarray}
\phi ^{1\ldots m} &=&\hat{U}_{1\ldots m}^{-1}\left( \Phi \otimes \Psi \right)
\notag \\
&=&\left( \hat{U}_{1\ldots k}^{-1}\Phi \right) \otimes \left( \hat{U}%
_{(k+1)\ldots m}^{-1}\Psi \right) ,
\end{eqnarray}
which contradicts the given condition that $\phi ^{1\ldots m}$ is fully
entangled. Hence the result is proven for entanglements as well as
separations.
\end{enumerate}

Because all partitions can be written as separation products of separations
and entanglements, we readily deduce that the result holds for all
partitions generally.

\

\begin{center}
{\large {$\mathbf{X.}$ \textbf{STATE PREPARATION}} }
\end{center}

In the stages paradigm, each outcome of a jump serves also as a preparation
or initial state for the next jump. This means that the dynamics of the
universe involves a sequence of \emph{ideal measurements }\cite{PERES:93}%
\emph{, }but\emph{\ }this does not mean that information is being extracted
by any external observer.

When we are discussing an actual physics experiment, the term \emph{state
preparation} will be reserved here to mean a particular class of stage jump,
from a state of the universe $\Psi _{n-1}$\ to a state of the universe $\Psi
_{n}$\ such that $\Psi _{n}$\ is of the separable form
\begin{equation}
\Psi _{n}\equiv \psi _{\lbrack 12\ldots P]}\otimes \Theta _{\lbrack
(P+1)\ldots N]},
\end{equation}
for some $P$ such that $1\leqslant P<N$, where $\psi _{\lbrack 12\ldots P]}$
is in $\mathcal{H}_{\left[ 12\ldots P\right] },\Theta _{\left[ (P+1)\ldots N%
\right] }$ is in $\mathcal{H}_{\left[ (P+1)\ldots N\right] },$\ and $%
\mathcal{H}_{\left[ 12\ldots P\right] }\otimes \mathcal{H}_{\left[
(P+1)\ldots N\right] }$ is a particular split of the total Hilbert space $%
\mathcal{H}_{\left[ 12\ldots N\right] }$. Without loss of generality, we may
always relabel the subregisters to give the above convenient representation.
More specifically, $\Psi _{n}$\ will be in the particular partition
\begin{equation}
\Psi _{n}\in \mathcal{H}_{\left[ 12\ldots P\right] }\bullet \mathcal{H}_{%
\left[ (P+1)\ldots N\right] }\equiv \mathcal{H}_{\left[ 12\ldots P\right]
\bullet \left[ (P+1)\ldots N\right] }  \label{200}
\end{equation}
of the total Hilbert space.

Depending on the split, $\psi _{\lbrack 12\ldots P]}$ may be thought of as
the state of the subject of the experiment whilst $\Theta _{\lbrack
(P+1)\ldots N]}$ is the state of the observer plus environment plus wider
universe. It is clear from our paradigm, however, that this is not an
intrinsic description; which is observer and which is subject will be
irrelevant except on emergent scales. In actual physics experiments, $P$\
could be relatively small, such as 1, or enormous, such as of the order $%
10^{100}$, but this would pale into relative insignificance given an $N$ of
the order $10^{182}$ or more. Under these circumstances, i.e. $1\leqslant
P\ll N$, it is reasonable to call $\psi _{\lbrack 12\ldots P]}$ the \emph{%
subject} (system under observation) and $\Theta _{\lbrack (P+1)\ldots N]}$\
the \emph{observer} (which includes the laboratory, the local environment
and the rest of the universe).

\

\begin{center}
{\large {$\mathbf{XI.}$ \textbf{TRANSITION AMPLITUDE FACTORS}} }
\end{center}

Before we can give a definition of what we mean by information exchange in
quantum systems, we need to make some observations concerning transition
amplitudes. In the following, it is the ``zipping'' properties of quantum
register inner products, discussed in $\S VI$, which combine with the
partition structure of the states concerned to produce the factorization
properties of transition amplitudes.

Let $\Psi _{n}$ and $\Psi _{n+1}$ be two successive states of the universe.
Each of these is a vector in $\mathcal{H}_{\left[ 1\ldots N\right]}$, the
total quantum register. Now from our discussion of entanglements and
separations, we know that $\mathcal{H}_{\left[ 1..N\right]}$ is the union of
a large number of partitions, the full set of which we call the natural
lattice of partitions and denoted by $\frak{L}\left( \mathcal{H}_{\left[ 1..N%
\right]}\right)$. Each element $\Psi $\ in $\mathcal{H}_{\left[ 1..N\right]}$
lies in a unique partition $P_{\Psi }$\ in $\frak{L}\left(\mathcal{H}_{\left[
1..N\right] }\right)$ and has $F\left[ P_{\Psi }\right]$ factors, each
factor lying in a different block associated with $P_{\Psi}$. For each state
$\Psi _{n}$, the pattern of the associated blocks defines a unique split $%
\mathcal{S}_{n}$ of the quantum register.

There are two cases to consider: $i)$ $\Psi _{n}$, $\Psi _{n+1}$ are in the
same partition and $ii)$ $\Psi _{n}$ and $\Psi _{n+1}$ are in different
partitions. These need to be discussed separately.

$i)$ Suppose that $\Psi _{1},\Psi _{2}\in \mathcal{H}_{\left[ 1\ldots N%
\right] }$ lie in the same partition, i.e., $P_{\Psi _{1}}=P_{\Psi _{2}}$,
and are not in the full entanglement $\mathcal{H}^{1\ldots N}$. Then each
state has at least two factors (assuming $N>2)$. Because they lie in the
same partition, they have the same number of factors, each of which is a
separation or an entanglement. Suppose there are $F$ such factors, such that
\begin{equation}
\Psi _{i}=\psi _{i}^{\left( 1\right) }\otimes \psi _{i}^{\left( 2\right)
}\otimes \ldots \otimes \psi _{i}^{\left( F\right) },\;\;\;i=n,n+1,
\end{equation}
where each factor $\psi _{i}^{\left( k\right) }$ belongs to the $k^{th}$
block in the partition $P_{\Psi _{i}}$. Then by virtue of the ``zipping''
properties of the subregisters, the inner product $(\Psi _{n+1},\Psi _{n})$
necessarily factorizes into the same number $F\equiv F\left[ P_{\Psi _{n}}%
\right] =F\left[ P_{\psi _{n+1}}\right] $ of factors formally, i.e.,
\begin{equation}
(\Psi _{n+1},\Psi _{n})=\left( \psi _{n+1}^{\left( 1\right) },\psi
_{n}^{\left( 1\right) }\right) \left( \psi _{n+1}^{\left( 2\right) },\psi
_{n}^{\left( 2\right) }\right) \ldots \left( \psi _{n+1}^{\left( F\right)
},\psi _{n}^{\left( F\right) }\right) .
\end{equation}
Each factor in this product is associated with a given factor in the split $%
\mathcal{S}_{n}$ of $\mathcal{H}_{\left[ 1\ldots N\right] }$.

$ii)$ Suppose on the other hand that $P_{\Psi _{n}}\neq P_{\Psi _{n+1}}$,
i.e., $\Psi _{n}$ and $\Psi _{n+1}$ belong to different partitions. Then by
inspection we find that for any transition involving such a change of
partition, the number of factors $F\left[ (\Psi _{n+1},\Psi _{n})\right] $
in the amplitude $(\Psi _{n+1},\Psi _{n})$\ satisfies the relation
\begin{equation}
F\left[ (\Psi _{n+1},\Psi _{n})\right] \leqslant \min \left\{ F\left( \Psi
_{n}\right) ,F\left( \Psi _{n+1}\right) \right\} <\max \left\{ F\left( \Psi
_{n}\right) ,F\left( \Psi _{n+1}\right) \right\} .
\end{equation}

The upper bound $\max \left\{ F\left( \Psi _{n}\right) ,F\left(
\Psi_{n+1}\right) \right\}$ is attained only if there is no partition
change. If either $\Psi _{n}$\ or $\Psi _{n+1}$\ is in the full entanglement
$\mathcal{H}^{1\ldots N}$, then $F\left[ (\Psi _{n+1},\Psi _{n})\right] $
attains its lowest bound, unity.

\begin{center}
\begin{figure}[t]
\centerline{\epsfxsize=5.0in \epsfbox{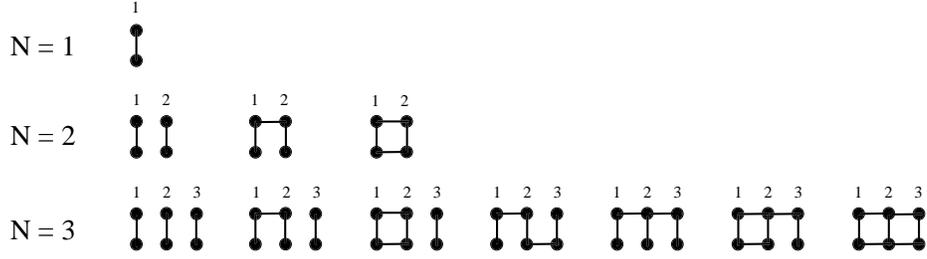}}
\caption{Topologically inequivalent quantum zip architecture for rank-$1$, $%
2 $ and $3$ quantum registers.}
\end{figure}
\end{center}

The set of all possible patterns of factorization of transition amplitudes
becomes increasingly more complex as the rank of the quantum register
increases. In Figure $1$, we show all topologically inequivalent quantum zip
diagrams for rank-$1$, $2$ and $3$ quantum registers, disregarding the
direction of time. Vertical lines represent inner products between
subregisters whilst horizontal lines represent entanglements.

Table 1 shows the number of formal factors $F$ in the amplitudes between all
the various possible initial and final state types for a rank-$3$ quantum
register:

\begin{center}
\begin{tabular}{c|ccccc}
$F$ & $\mathrm{\psi }_{123}$ & $\mathrm{\psi }_{1}^{23}$ & $\mathrm{\psi }%
_{2}^{13}$ & $\mathrm{\psi }_{3}^{12}$ & $\mathrm{\psi }^{123}$ \\ \hline
$\mathrm{\phi }_{123}$ & $\mathrm{3}$ & $\mathrm{2}$ & $\mathrm{2}$ & $%
\mathrm{2}$ & $\mathrm{1}$ \\
$\mathrm{\phi }_{1}^{23}$ & $\mathrm{2}$ & $\mathrm{2}$ & $\mathrm{1}$ & $%
\mathrm{1}$ & $1$ \\
$\mathrm{\phi }_{2}^{13}$ & $\mathrm{2}$ & $\mathrm{1}$ & $\mathrm{2}$ & $%
\mathrm{1}$ & $1$ \\
$\mathrm{\phi }_{3}^{12}$ & $\mathrm{2}$ & $\mathrm{1}$ & $\mathrm{1}$ & $%
\mathrm{2}$ & $1$ \\
$\mathrm{\phi }^{123}$ & $\mathrm{1}$ & $1$ & $1$ & $1$ & $1$%
\end{tabular}

\

Table $1.$ The number of amplitude factors $F$ for given initial and final
states.
\end{center}

\

\begin{center}
{\large {$\mathbf{XII.}$ \textbf{THE ISOLATED QUANTUM SYSTEM}} }
\end{center}

In this section we discuss what can be said about an isolated quantum system
$S$, called the \emph{subject}, when it is described by a rank-$1$ quantum
register of dimension $d$. States of such a system will be assumed never to
factorize. There are three scenarios which we shall consider in turn. The
first two invoke the usual exophysical principles of standard quantum
mechanics whilst the third gives an endophysical account.

\

\begin{center}
\textbf{A. Complete isolation from exophysical observers}
\end{center}

In this scenario, exophysical observers stand outside of the subject $S$
with no information exchange whatsoever between it and them after it has
been prepared by them in a given state $\psi _{0}$ at initial exophysical
time $n=0$. The external observers have only two roles in this scenario:
first they prepare the subject state and then they arrange to keep it
isolated from the rest of the universe. They retain a knowledge of how the
subject state was prepared and are motivated to keep it isolated. We shall
argue that after preparation, the subject will behave in effect as if it
were an isolated subuniverse and, because it is not observed, it may be
considered by the observers to be ``frozen'' in time, even if it evolves
autonomically.

The issue here rests on what ``complete isolation'' means. Taken literally,
it can only mean that no test organized by the external observers is
performed on the subject after state preparation. No test means no outcome,
which means the state remains unchanged.

This is of course a rather trivial conclusion. The discussion is not quite
complete however, because we have to examine the possibility that the
subject might test itself and jump into a new state without reference to the
external observers. The question is then, what could the external observers
say under those circumstances?

The problem faced by the external observers in discussing this possibility
is analogous to the problem of parallel transport in general relativity.
This arises because tangent vector spaces at different points in a manifold
are distinct spaces. In principle, there is no a priori or natural way of
ensuring that bases in different tangent vector spaces coincide and in a
sense, such a concept has no direct physical meaning anyway.

It was because of this problem that the concepts of Lie differentiation and
covariant differentiation had to be devised. These provide a way of relating
bases in different spaces to each other. For example, in the case of a
Riemannian manifold, the metric over the manifold can be used to construct a
metric connection, which can then be used to determine how basis vectors
change as we move over the manifold. In our case, we need to define
carefully what we mean by the idea that the state of the system ``changes''
from $\psi _{n}$ to state $\psi _{n+1}$ at exotime $n+1$ in the absence of
any measurement of $\psi _{n+1}$ by the external observers.

To help us in the formal analysis and by analogy with the parallel transport
problem, we shall assume that states of the subject $S$ at different times
lie in different Hilbert spaces, i.e., we shall suppose that for $n\geqslant
0$, $\psi _{n}\in \mathcal{S}_{n},$ where the $\mathcal{S}_{n}$ are copies
of $\mathcal{S}$.

Now assuming that the stages paradigm holds, the observers know some things
for certain but have only a partial knowledge of other things. What they do
know is this: because they prepared the subject state and retained a
knowledge of it, the observers know all about the initial preferred basis
set $\mathsf{B}_{0}\equiv \left\{ |i,0\rangle :1\leqslant i\leqslant
d\right\} \in \frak{B(}\mathcal{S}_{0})$. Also, they know which particular
element of this basis set the prepared initial state $\psi _{0}$ happened to
be.

Believing that the subject may have evolved via its own internal dynamics by
some arbitrary number of jumps, the observers are entitled to assume that at
any moment $n>0$ of exotime after preparation, $\psi _{n}$ is some element
of another preferred basis $\mathsf{B}_{n}\equiv \left\{ |i,n\rangle
:1\leqslant i\leqslant d\right\} \in \frak{B(}\mathcal{S}_{n})$. However,
although they may believe that these things ``exist'', the observers have no
knowledge of either the $\mathsf{B}_{n}$ or the actual outcomes $\psi _{n}.$

Even with such limited information, the observers may always relabel the
elements in $\mathsf{B}_{0}$ so that $\psi _{0}=|1,0\rangle $. They can also
assume a formal relabelling for the unknown bases $\mathsf{B}_{n}$ such that
for each $n$, $\psi _{n}=|1,n\rangle $, because this does not actually
invoke any new knowledge. The question now is, what grounds are there for
relating the elements of the basis sets $\mathsf{B}_{0}$ and $\mathsf{B}_{n}$%
? They are bases for different copies of the same Hilbert space and
therefore something analogous to a connection would be required in order to
allow us to compare vectors in one copy with vectors in another.

To formally define a process of ``parallel transport of basis'', we
introduce a linear map $\hat{U}\left( \mathsf{B}_{n+1},\mathsf{B}_{n}\right)$
from $\mathcal{S}_{n}$ to $\mathcal{S}_{n+1}$ given by the expression
\begin{equation}
\hat{U}\left( \mathsf{B}_{n+1},\mathsf{B}_{n}\right) \equiv
\sum_{i=1}^{d}|i,n+1\rangle \langle i,n|.
\end{equation}
This map transports ket states in $\mathcal{S}_{n}$ into ket states in $%
\mathcal{S}_{n+1}$ and has properties associated with Rota incidence
algebras \cite{RAPTIS-99}. Such algebras encode some of the properties of
causal sets. For instance, we have the product rule
\begin{equation}
\hat{U}\left( \mathsf{B}_{n+2},\mathsf{B}_{n+1}\right) \hat{U}\left( \mathsf{%
B}_{n+1},\mathsf{B}_{n}\right) =\hat{U}\left( \mathsf{B}_{n+2},\mathsf{B}%
_{n}\right) ,
\end{equation}
but the ``product'' $\hat{U}\left( \mathsf{B}_{n+1},\mathsf{B}_{n}\right)
\hat{U}\left( \mathsf{B}_{n+2},\mathsf{B}_{n+1}\right) $ is not defined.
There has to be some chain of causality for such products to be meaningful.

The map $\hat{U}\left( \mathsf{B}_{n+1},\mathsf{B}_{n}\right) $ is
invertible, with inverse map
\begin{equation}
\hat{U}^{-1}\left( \mathsf{B}_{n+1},\mathsf{B}_{n}\right) \equiv \hat{U}%
\left( \mathsf{B}_{n},\mathsf{B}_{n+1}\right) =\sum_{i=1}^{d}|i,n\rangle
\langle i,n+1|.
\end{equation}
Then we have the results
\begin{equation}
\hat{U}^{-1}\left( \mathsf{B}_{n+1},\mathsf{B}_{n}\right) \hat{U}\left(
\mathsf{B}_{n+1},\mathsf{B}_{n}\right) =\hat{I}_{n},\;\;\;\hat{U}\left(
\mathsf{B}_{n+1},\mathsf{B}_{n}\right) \hat{U}^{-1}\left( \mathsf{B}_{n+1},%
\mathsf{B}_{n}\right) =\hat{I}_{n+1},
\end{equation}
where $\hat{I}_{n}$ and $\hat{I}_{n+1}$ are the identity operators in $%
\mathcal{S}_{n}$ and $\mathcal{S}_{n+1}$ respectively. The inverse map is
formally equivalent to the ``adjoint'' map, i.e.,
\begin{equation}
\hat{U}^{-1}\left( \mathsf{B}_{n+1},\mathsf{B}_{n}\right) =\hat{U}^{+}\left(
\mathsf{B}_{n},\mathsf{B}_{n+1}\right) \equiv \sum_{i=1}^{d}|i,n\rangle
\langle i,n+1|,
\end{equation}
which takes states in the dual $\left( \text{bra}\right) $ space $\mathcal{S}%
_{n}^{\ast }$ into states in the dual space $\mathcal{S}_{n+1}^{\ast }$.

If now we apply the map $\hat{U}\left( \mathsf{B}_{n},\mathsf{B}_{0}\right) $
to the state $\psi _{0}$ we find
\begin{equation}
\hat{U}\left( \mathsf{B}_{n},\mathsf{B}_{0}\right) \psi _{0}=\psi _{n}\text{.%
}
\end{equation}

The effect of this is to make the internal jump of the subject state look
formally like the result of some unitary evolution, such as that given by
the integration of Schr\"{o}dinger evolution in continuous time. We may now
formally ``undo'' this evolution by transforming to the Heisenberg picture,
effectively rotating the basis $\mathsf{B}_{n}$ into $\mathsf{B}_{0}$. Now
precisely because the external observers have no interaction with $S$ after
state preparation, this transformation has no physical consequences for
them; there is no measurement undertaken by the external observers
subsequent to state preparation which could register the effect of the
change of picture.

Our conclusion is, therefore, that as long as the subject remains isolated,
the external observers are entitled to regard it as frozen in time, even if
they had grounds for believing it had some autonomous evolution.

\

\begin{center}
\textbf{B. Standard exophysical quantum description}
\end{center}

The above discussion is close to triviality because it ignores the physical
presence of the external observers after state preparation. It is, in
effect, a solipsist view of a subuniverse in which only the subject exists.
In this subsection we discuss the conventional quantum description of an
experiment on a nominally isolated quantum subject system in the active
presence of external observers. Now the external observers have three roles:
$i)$ they prepare the subject state at time $n=0$, $ii)$ they isolate the
subject during some interval $\left( 0,M\right) $, of their endotime, where
typically $M\gg 0,$ and finally, $iii)$ they test the subject state at time $%
M$.

The analysis proceeds as for the first scenario, except for the final phase,
state testing, which occurs at time $M.$ Now the external observers play an
active role, because they create the testing equipment. This defines a final
preferred basis $\mathsf{B}_{M},$ which is therefore now known to the
observers. The observers also retain a knowledge of the initial preferred
basis $\mathsf{B}_{0}$ and indeed of the initial state $\psi _{0}\in \mathsf{%
B}_{0}$. It is the simultaneous knowledge of both bases which cannot be
eliminated by any passive rotation of basis. A rotation of one basis must
also be applied to the other basis, so that transition amplitudes between
elements of the two bases remain invariant and cannot be eliminated.

Another important difference between this scenario and the previous one is
that, \emph{before} the final test, the principles of quantum mechanics do
not now permit the observers to assume that the final state is any
particular element of the final preferred basis $\mathsf{B}_{M},$ such as $%
\psi _{M}=|1,M\rangle $. Therefore, the Rota incidence algebra discussion
can only be undertaken \emph{after} the whole experiment is over and outcome
$\psi _{M}$ has been observed. By that time, however, it will no longer be
reasonable for the observers to argue that the subject state has not
changed, because a real active change in it will have been registered for
certain by their equipment.

In standard quantum mechanics, external observers are a crucial component of
information exchange. It is their knowledge of \emph{both} the initial and
final bases which cannot be transformed away by any unitary transformation
of basis. This is why either the Schr\"{o}dinger picture or Heisenberg
picture can be used in standard quantum mechanics. This is not the case in
any paradigm which has no external observers and no state reduction.

\

\begin{center}
\textbf{C. The endophysical description}
\end{center}

The second scenario above is exophysical and is therefore one we wish to
replace by an endophysical discussion. We now discuss what happens from the
endophysical perspective, when the ``observers'' are contained within a
greater quantum system $U$ (the universe), of which the original subject $S$
is but a part. The discussion in this situation requires more care
concerning the tests involved.

Now regardless of whether an exophysical or endophysical description is
being used, a quantum experiment generally requires a sequence of three
things to happen. First the subject has to be prepared in some initial state
(state preparation). Then the initial state then has to be given time to
evolve in isolation. Finally, at the end of each run of the experiment, the
state has to be tested. In the real world of experimental physics, each of
these steps will be very complex, involving extremely sophisticated
equipment and experimental protocols, even if the standard quantum
description appears straightforward. Let us consider each of these steps
separately for a given run of the experiment.

\begin{center}
$\mathbf{i)}$\textbf{\ State preparation}
\end{center}

Assuming that the greater quantum system $U$ containing the observer and the
subject can be described via the stages paradigm, state preparation means
that the state $\Psi _{n}$ of the universe at initial time $n=0$ is of the
form
\begin{equation}
\Psi _{0}\equiv \Theta _{0}\otimes \psi _{0},  \label{888}
\end{equation}
where the pure state $\Theta _{0}$ (the \emph{complement}) represents the
observers, apparatus and environment external to the subject, which is
represented by $\psi _{0}$ as before. Note that $\Theta _{0}$ will
necessarily be different for each run of the experiment, because the
universe as a whole is not reversible. On the other hand, the subject state $%
\psi _{0}$ can be assumed to be identical at the start of each run. If this
last condition is relaxed, then a density matrix discussion of the
experiment will be required.

The complement $\Theta _{0}$ is an element of some large rank quantum
register $\mathcal{A}_{0}$ whilst $\psi _{0}$ is an element of $\mathcal{S}%
_{0}$, as before. The initial state of the universe $\Psi _{0}$ is an
element of the quantum register $\mathcal{H}_{0}$ $\equiv \mathcal{A}%
_{0}\otimes \mathcal{S}_{0}$. Although $\mathcal{H}_{0}$ contains both
entangled and separable states, state preparation means that $\Psi _{0}$ is
of the separable form $\left( \ref{888}\right) $ and we are entitled to
write
\begin{equation}
\Psi _{0}\in \mathcal{A}_{0}\bullet \mathcal{S}_{0}\subset \mathcal{A}%
_{0}\otimes \mathcal{S}_{0}.
\end{equation}
In other words, state preparation in a physics experiment is equivalent to
ensuring that an initial state of the universe is an element of a partition
with two or more factors (blocks).

\begin{center}
$\mathbf{ii)}$\textbf{\ Isolation}
\end{center}

Now in a real experiment, the observers generally contrive in some way to
ensure that the system under investigation remains isolated for a reasonable
length of their time. Therefore, in our endophysical description of such a
process, we may suppose that there is a sequence of jumps of the state of
the universe,
\begin{equation}
\Psi _{0}\rightarrow \Psi _{1}\rightarrow \ldots \rightarrow \Psi
_{n}\rightarrow \ldots \rightarrow \Psi _{M-1},\;\;\;M\gg 0.
\end{equation}
during which the subject state remains isolated, i.e., we have
\begin{equation}
\Psi _{n}\equiv \Theta _{n}\otimes \psi _{n}\in \mathcal{A}_{n}\bullet
\mathcal{S}_{n}\subset \mathcal{A}_{n}\otimes \mathcal{S}_{n},\;\;\;0%
\leqslant n<M,  \label{777}
\end{equation}
where $\mathcal{A}_{n}$ is a copy of $\mathcal{A}_{0}$, provided we have not
entered the partition change regime (which is where we state that real
information exchange occurs).

In real physics experiments, the observers do not in general have absolute
control of the universe and total isolation of a subject system cannot be
guaranteed in general. For example, in particle scattering experiments,
there is always the possibility of background effects, such as high energy
cosmic rays passing through the subject system, interfering with any
particular run of the experiment. It is only \emph{after} the scattering
information has been acquired that the observers can deduce what level of
isolation had in fact been achieved.

From the stages paradigm point of view, complete isolation can be guaranteed
by virtue of Theorem $3,$ \emph{if} each test $\hat{\Sigma}_{n}$ of the
universe between state preparation and outcome definitely factorizes into
the form
\begin{equation}
\hat{\Sigma}_{n}=\hat{A}_{n}\otimes \hat{S}_{n},  \label{123}
\end{equation}
where $\hat{A}_{n}\in \mathbb{S}\left( \mathcal{A}_{n}\right) $ and $\hat{S}%
_{n}\in \mathbb{H}\left( \mathcal{S}_{n}\right) .$ Background effects cannot
occur if this factorization condition holds. Another way of understanding
background effects is that they involve entanglement between observer and
subject.

Given perfect isolation, Theorem 3 tells us that because $\hat{\Sigma}_{n}$
is strong (according to our principles), then its preferred basis $\mathsf{B}%
_{n}$ is completely factorizable, i.e., is of type $\left( 0,\dim \mathcal{A}%
\times \dim \mathcal{S}\right) $. Indeed, we may write
\begin{equation}
\mathsf{B}_{n}=\mathsf{B}_{A_{n}}\bullet \mathsf{B}_{S_{n}},
\end{equation}
where $\mathsf{B}_{A_{n}}$ is the preferred basis for $\hat{A}_{n}$ and $%
\mathsf{B}_{S_{n}}$ is the preferred basis for $\hat{S}_{n}$. We can now
apply the Rota incidence algebra discussion to each subject state basis $%
\mathsf{B}_{S_{n}}$ so that it is rotated back into the initial preferred
basis $\mathsf{B}_{S_{0}}$. The incidence algebra operators are extended to
the universal register in a straightforward way, i.e., for each $n$ between $%
0$ and $M$ we define
\begin{equation}
\hat{U}\left( \mathsf{B}_{n},\mathsf{B}_{0}\right) \equiv {\hat{I}}%
_{A_{n}}\otimes \sum_{i=1}^{d}|i,n\rangle \langle i,0|,\;\;\;0\leqslant n<M,
\end{equation}
where ${\hat{I}}_{A_{n}}$ is the identity over $\mathcal{A}_{n}$. Then
\begin{equation}
\Psi _{n}\rightarrow \Psi _{n}^{\prime }\equiv \hat{U}^{-1}\left( \mathsf{B}%
_{n},\mathsf{B}_{0}\right) \Psi _{n}=\Theta _{n}\otimes |1,0\rangle =\Theta
_{n}\otimes \psi _{0}.
\end{equation}
The result is that during the period of isolation, the state of the subject
system may be regarded as subject to what we call a \emph{null test }\cite
{JAROSZKIEWICZ-01A}, the principal characteristic of which being that it
tests one of its eigenstates, which therefore passes through unchanged.

These Rota incidence algebra transformations are equivalent to transforming
to a partial Heisenberg picture wherein the subject state is frozen but the
rest of the universe is not. Because in principle there is no formal
difference in the stages paradigm between subject and complement, it should
be possible to exchange roles and freeze the observers but not the subject.
However, we should keep in mind that in real situations, the register $%
\mathcal{A}$ associated with the observers will have vastly greater rank
than the register $\mathcal{S}$ associated with the subject. It is the
vastly greater complexity associated with $\mathcal{A}$ which permits us to
use anthropomorphic terminology occasionally, as if the observers had some
sort of choice in what they were doing. In fact, the entire system $U$ is
behaving simply as a quantum automaton.

From now on, we shall drop the formal distinction between different temporal
copies of the same vector space. We introduced this notion in order to
facilitate the Rota algebra discussion. In fact, the concept of different
copies of spaces corresponding to different times is a block universe idea
which has no place in process time thinking. Therefore, both $\Psi _{n}$ and
$\Psi _{n}^{\prime }$ may be considered to be in the same quantum register.
An important point about the Rota analysis is that the observers believe
that only one of these vectors represents the true state of the subject at
time $n$.

The sequence of states of the universe during isolation is now given by
\begin{equation}
\Theta _{0}\otimes \psi _{0}\rightarrow \Theta _{1}\otimes \psi
_{0}\rightarrow \ldots \rightarrow \Theta _{n}\otimes \psi _{0}\rightarrow
\ldots \rightarrow \Theta _{M-1}\otimes \psi _{0}.
\end{equation}
Essentially, the subject state can be regarded as frozen during the period
of isolation, simply because it not being observed, whilst the rest of the
universe undergoes dynamical change.

\

\begin{center}
$\mathbf{iii)}$\textbf{\ Test and outcome}
\end{center}

The most important aspect of the discussion concerns the nature of the test $%
\hat{\Sigma}_{M}$ at the end of the period of isolation and the nature of
the tests which follow it. Test $\Sigma _{M}$ will still be separable and of
the specific form (\ref{123}), but the Rota incidence algebra transformation
is not applied for the final time $M$. This is because the act of
measurement means that the observers believe that the hitherto isolated
subject state has made an active jump into some eigenstate of their
measuring equipment.

The final test $\Sigma _{M}$ will have a preferred basis $\mathsf{B}_{M}=%
\mathsf{B}_{A_{M}}\bullet \mathsf{B}_{S_{M}}$, where $\mathsf{B}%
_{A_{M}}\equiv \left\{ \Theta _{M}^{\alpha }:1\leqslant \alpha \leqslant
\dim \mathcal{A}\right\} $ and $\mathsf{B}_{S_{M}}\equiv \left\{ \phi
_{M}^{\beta }:1\leqslant \beta \leqslant \dim \mathcal{S}\right\} $, and so
the state of the universe at time $M$ is of the form $\Psi _{M}=\Theta
_{M}^{\alpha }\otimes \phi _{M}^{\beta }$ for some specific value of $\alpha
$ and of $\beta .$

Tests after time $M$ cannot take the form (\ref{123}). Otherwise Theorem $3$
tells us that isolation was continuing beyond time $n=M$. Therefore, in
order for information to be exchanged between the subject and the complement
after time $M$, $\hat{\Sigma}_{M+1}$ must be basis inequivalent to $\hat{%
\Sigma}_{M}$. In other words, information exchange requires a change of
partition associated with the test of the universe. There is no way around
this conclusion.

Once the universe has jumped into stage $\Omega _{M},$ the stages paradigm
dynamical principles take over. The information content $I_{M}$ now contains
information about the final state $\phi _{M}^{\beta }$ of the subject. The
rules $\mathcal{R}_{M}$ now take over and dictate that the subsequent test $%
\hat{\Sigma}_{M+1}$ will depend on this particular outcome. This particular
outcome essentially influences the subsequent history of the universe in
such a way that the information that the subject state jumped into state $%
\phi _{M}^{\beta }$ is encoded into all subsequent stages (at least for as
long as the observers retain a memory of the outcomes of the experiment).

After many runs of the same experiment, a frequency distribution for the
various outcomes $\phi _{M}^{\beta _{1}},\phi _{M}^{\beta _{1}},\ldots $
will be encoded (registered) into the state of the universe and eventually
the observers can compare this with the theoretical probability $P\left(
\phi _{M}^{\alpha }|\psi _{0}\right) $ of the transition $\psi
_{0}\rightarrow \phi _{M}^{\alpha }$. This is given in self-explanatory
notation by the rule
\begin{eqnarray}
P\left( \phi _{M}^{\alpha }|\psi _{0}\right) &=&\sum_{\Theta
_{1}}\sum_{\Theta _{2}}\ldots \sum_{\Theta _{M}}P\left( \Theta _{M}\otimes
\phi _{M}^{\alpha }|\Theta _{M-1}\otimes \psi _{0}\right) \times  \notag \\
&&\;\;\;P\left( \Theta _{M-1}\otimes \psi _{0}|\Theta _{M-2}\otimes \psi
_{0}\right) \ldots P\left( \Theta _{1}\otimes \psi _{0}|\Theta _{0}\otimes
\psi _{0}\right)  \notag \\
&=&\sum_{\Theta _{1}}\sum_{\Theta _{2}}\ldots \sum_{\Theta _{M}}P\left(
\Theta _{M}|\Theta _{M-1}\right) |\langle \phi _{M}^{\alpha }|\psi
_{0}\rangle |^{2}\times  \notag \\
&&\;\;\;P\left( \Theta _{M-1}|\Theta _{M-2}\right) \ldots P\left( \Theta
_{1}|\Theta _{0}\right)  \notag \\
&=&|\langle \phi _{M}^{\alpha }|\psi _{0}\rangle |^{2}\sum_{\Theta
_{1}}\sum_{\Theta _{2}}\ldots \sum_{\Theta _{M}}\prod_{n=1}^{M}|\langle
\Theta _{n}|\Theta _{n-1}\rangle |^{2}  \notag \\
&=&|\langle \phi _{M}^{\alpha }|\psi _{0}\rangle |^{2},
\end{eqnarray}
which is the standard quantum result. In this calculation, summation is over
all the basis elements of the bases for intermediate stages. We note that
there are no interference terms.

This calculation does not take into account the fact that the state $\Theta
_{0}$ of the complement at the start of each run would be different each
time (necessarily so, because after each run, the universe has registered
new information). Neither does it take into account the probability
distribution of the tests $\hat{A}_{n}$ of the complement during isolation.
It is not hard to see that these effects would not alter the conclusion in
any way. In other words, quantum experiments on isolated systems can be
undertaken and the rules of quantum mechanics can be applied to those
isolated systems, even when the observers are themselves evolving according
to the rules of quantum mechanics.

\

\begin{center}
{\large {$\mathbf{XIII.}$ \textbf{HIGHER RANK SUBJECT SYSTEMS}} }
\end{center}

For completeness, we discuss now what may happen when the subject system
consists of two or more subregisters. Now we are able to discuss separations
and entanglements of the subject state, which gives some important
constraints on our ability to ``parallel-transport away'' changes in state
during isolation. For simplicity we shall restrict our attention to the case
when the subject register $\mathcal{S}$ consists of two subregisters, i.e., $%
\mathcal{S=}$ $\mathcal{H}_{\left[ 12\right] }\equiv \mathcal{H}_{1}\mathcal{%
\otimes H}_{2}.$

There are three cases to consider during the period of isolation; in each
case we restrict our attention to the subject register.

\

\begin{center}
\noindent $\mathbf{i)}$ \textbf{separable to separable:}
\end{center}

Consider a jump of the subject state of the form $\psi _{12}\rightarrow \phi
_{12}$, where each of these states is in the separation $\mathcal{H}_{12}$.
With our conventions for separations and entanglements, we can rewrite this
process in a way reminiscent of consistent histories:
\begin{eqnarray}
\psi _{12}\rightarrow \phi _{12} &\Rightarrow &\psi _{1}\otimes \psi
_{2}\rightarrow \phi _{1}\otimes \phi _{2}  \notag \\
&=&\left( \psi _{1}\rightarrow \phi _{1}\right) \otimes \left( \psi
_{2}\rightarrow \phi _{2}\right) ,
\end{eqnarray}
where the tensor product in the last term carries a somewhat different
meaning to that hitherto. It is a sort of product of ``histories''.
Essentially, the dynamical evolution here suggests that there are two
completely distinct, noninteracting subject systems, each of which appears
to be evolving in its own subuniverse. Moreover, each of these subuniverses
is no different in its properties to the rank-1 subject register discussed
extensively in the previous section. We can see then, that in this
particular case, a passive local unitary transformation applied to the
quantum register can transform away the change in the state, exactly as in
the Rota algebra discussion applied above.

\

\begin{center}
\noindent $\mathbf{ii)}$ \textbf{separable to entangled (and vice-versa)}
\end{center}

With a jump of the form $\psi _{12}\rightarrow \phi ^{12}$ there is a change
of partition, from a state in the separation $\mathcal{H}_{12}$ to a state
in the entanglement $\mathcal{H}^{12}$. It will be seen upon inspection that
there is no way of performing any \emph{local }unitary transformation which
can transform away such a change in the state. The reason of course is
directly associated with Theorem $4$, which states that partitions are
invariant to local unitary transformations. This sort of jump therefore
represents a nontrivial internal change of the subject state involving both
subregisters.

If the observers external to the subject wish to use the Rota algebra method
to transform away the change of the subject, then they can only do so if
they perform a non-local unitary transformation. In other words, they have
to ignore the fact that $\mathcal{S}$ is a tensor product space.

\

\begin{center}
\noindent $\mathbf{iii)}$ \textbf{entangled to entangled}
\end{center}

A jump of the form $\psi ^{12}\rightarrow \phi ^{12}$ raises the interesting
mathematical question of whether we can always find some \emph{local}
unitary transformation which can transform one entangled state $\psi ^{12}$
into any other entangled state $\phi ^{12}$. If the answer were \emph{yes},
then taking into account our previous analysis, we would conclude that
without a change of partition of the subject state, we could always maintain
the fiction that there is no intrinsic quantum dynamics (i.e., we could
always regard fixed-partition change as due to a passive transformation of
basis).

To answer this question we shall consider a two qubit quantum register.
Suppose $\psi ^{12}$ and $\phi ^{12}$ are two states in the entanglement $%
\mathcal{H}^{12}$ of the register $\mathcal{H}_{1}\otimes \mathcal{H}_{2}$.
According to Schmidt decomposition, we can always find a decomposition of
each state in the following form:
\begin{eqnarray}
\psi ^{12} &=&\sqrt{p}|a\rangle _{1}\otimes |b\rangle _{2}+\sqrt{1-p}|\bar{a}%
\rangle _{1}\otimes |\bar{b}\rangle _{2},  \notag \\
\phi ^{12} &=&\sqrt{q}|u\rangle _{1}\otimes |v\rangle _{2}+\sqrt{1-q}|\bar{u}%
\rangle _{1}\otimes |\bar{v}\rangle _{2},
\end{eqnarray}
where $0\leqslant p,q\leqslant {\frac{1}{2}}$, \textsf{B}$_{1}^{a}\equiv
\left\{ |a\rangle _{1},|\bar{a}\rangle _{1}\right\} $ and \textsf{B}$%
_{1}^{u}\equiv \left\{ |u\rangle _{1},|\bar{u}\rangle _{1}\right\} $ are
orthonormal bases for $\mathcal{H}_{1}$ and \textsf{B}$_{2}^{b}\equiv
\left\{ |b\rangle _{2},|\bar{b}\rangle _{2}\right\} $ and \textsf{B}$%
_{2}^{v}\equiv \left\{ |v\rangle _{2},|\bar{v}\rangle _{2}\right\} $ are
orthonormal bases for $\mathcal{H}_{2}.$ Here, the real numbers $p$ and $q$
can be interpreted as conditional probabilities.

Now it is always possible to construct a unitary transformation $\hat{U}_{1}$
which transforms \textsf{B}$_{1}^{a}$ into \textsf{B}$_{1}^{u}$, such that
\begin{equation}
\hat{U}_{1}|a\rangle _{1}=|u\rangle _{1},\;\;\;\hat{U}_{1}|\bar{a}\rangle
_{1}=\hat{U}_{1}|\bar{u}\rangle _{1}.
\end{equation}
In fact, $\hat{U}_{1}$ is unique and given by
\begin{equation}
\hat{U}_{1}=|u\rangle _{1}\langle a|_{1}+|\bar{u}\rangle _{1}\langle \bar{a}%
|_{1}.
\end{equation}
Likewise there exists a unique unitary transformation $\hat{V}_{2}$ which
transforms \textsf{B}$_{2}^{b}$ into \textsf{B}$_{2}^{v}$ such that
\begin{equation}
\hat{V}_{2}|b\rangle _{2}=|v\rangle _{2},\;\;\;\hat{V}_{2}|\bar{b}\rangle
_{2}=\hat{V}_{2}|\bar{v}\rangle _{2}.
\end{equation}

The tensor product $\hat{W}_{12}\equiv \hat{U}_{1}\otimes \hat{V}_{2}$ is a
local unitary transformation on $\mathcal{H}_{\left[ 12\right] }\equiv
\mathcal{H}_{1}\otimes \mathcal{H}_{2}$ which has the specific effect of
transforming $\psi ^{12}$ into a near clone of $\phi ^{12}$:
\begin{equation}
\psi ^{12}\rightarrow \hat{W}_{12}\psi ^{12}=\sqrt{p}|u\rangle _{1}\otimes
|v\rangle _{2}+\sqrt{1-p}|\bar{u}\rangle _{1}\otimes |\bar{v}\rangle _{2}.
\end{equation}
However, for $p\neq q$ it is clear that there is no way that we could
transform $\psi ^{12}$ into $\phi ^{12}$ exactly via any local unitary
transformation.

The conclusion from this is that \emph{if }\ an entangled factor of a state
of the subject jumps into another entangled factor within the same
entanglement, the change in that factor cannot always be interpreted in
terms of a passive, local unitary transformation of basis. In other words,
real dynamical changes can occur \emph{within} an entanglement, not
involving any change of partition.

This is an important result. It means that a universal quantum register can
be discussed in terms of isolated subsystems evolving within greater
subsystems. We can imagine a Schr\"{o}dinger's cat experiment locked away
within a box, such that even though we have no contact with the contents of
the box, we can legitimately imagine it evolving dynamically, with state
reduction taking place out of sight inside the box. Note however, on the
basis of our analysis in the previous section, that even though real
dynamical changes may occur within a subject state, no consequences of those
changes can be communicated to the observers unless there is a change in
partition involving the observers and the subject, i.e., effectively
entangling their states. In the case of the Schr\"{o}dinger's cat
experiment, this corresponds to opening the box. Essentially, states
evolving within fixed partitions behave as if they were in separate
subuniverses, rather like regions of spacetime divided by event horizons,
for as long as those partitions persist.

\

\begin{center}
{\large {$\mathbf{XIV.}$ \textbf{THE BASIC PRINCIPLE OF ENDOPHYSICAL
INFORMATION EXCHANGE}} }
\end{center}

By considering these basic examples we can see the appearance of a criteria
for defining real (i.e., intrinsic) dynamical changes in quantum states, as
opposed to those which could be removed by passive transformations. This
leads us to state what we mean by endophysical information transfer:

\

``\emph{any quantum process in which meaningful information is exchanged
within a quantum system is always accompanied by a change in partition''}.

\

We shall call this the principle of endophysical information exchange,
because it does not rely on any notion of observer or system \emph{per se}.
It says what information exchange means for closed systems.

Whilst this criterion does not at first sight look much like the
conventional picture of products of states representing observers and
systems changing in time due to interactions, it has a number of features
which are physically appealing. First, it relies on the existence of an
underlying quantum register. We have already demonstrated that the stage
paradigm based on these leads to causal set structures \cite
{JAROSZKIEWICZ-03B}. Our concept of information exchange is a natural one in
this paradigm.

Second, there is no need to introduce the concepts of \emph{systems} or
\emph{observers} to define information exchange. Changes of partition
involve mathematical relationships which apply equally well to either
concept. This therefore gives a truly endophysical picture of quantum system
dynamics. Certainly, if we needed to, we could always associate large
numbers of subregisters in a given block (factor of a partition) with an
``observer'', but this need not have any intrinsic meaning. It would be
whatever happened to the subsequent tests and outcomes which would determine
the viability of any defined ``observer'' or system concept. One criterion
for this would be the relative persistence in exotime of various patterns of
factorization. This is motivated by the observation that, on typical
macroscopic scales associated with humans, we do not normally see
macroscopic objects suddenly disappearing. All structures do disappear
eventually, however, given enough jumps.

Third, when there is no change in partition over a number of jumps, then
essentially the various blocks in the partition behave much like isolated
subjects and observers between which no real information exchange occurs. It
is only when partitions change that we can be sure that real dynamical
exchange between systems and observers have occurred. When these occur in
real experiments, changes of partition which may start off involving a
relatively small number of subregisters may be amplified enormously over
time, so that the effect in the emergent limit will be equivalent to the
conventional picture of semiclassical observers appearing to record changes
in quantum subsystems. With large rank quantum registers, it should be
possible to represent memory and data storage within the complex of
partition structure available. That cannot be the entire story however; it
is the tests on states which will determine in what sense partition
structure represents memory and data. In other words, the whole process of
quantum measurement involves the entire dynamics, which must include the
states and the tests. Indeed, how tests are determined is as fundamental to
the running of the universe as what the outcomes of those tests are.

A fourth point about our definition of information exchange is that
partition change is inherently consistent with the fundamental principle in
quantum mechanics that the acquisition of real information from a state
necessarily destroys that state. In fact, a partition change necessarily
destroys all factor states involved, so that in essence, we can say that a
quantum measurement changes not only the system being observed but the
observer as well.

\

\begin{center}
{\large {$\mathbf{XV.}$ \textbf{CONCLUDING REMARKS}} }
\end{center}

The basic idea we have put forward is remarkably simple but has many
implications. Many aspects of the idea remain to be developed. We have had
only a limited opportunity here to discuss the details of the sort of
experiments actually performed in physics laboratories. Although the
principles are universal, such experiments are quite atypical of the tests
of the universe occurring naturally. A physics experiment generally involves
vast numbers of subregisters, most of which will be associated with the
apparatus and the immediate environment, not to mention the physicists
running the experiment. Such experiments need many stages before completion,
but from our point of view are no more than enormous irreversible
amplifications of elementary partition changes.

The implications of these concepts will be reported in due course.

\

\begin{center}
{\large ACKNOWLEDGMENTS }
\end{center}

J.E. acknowledges the support of the UK E.P.S.R.C.

\

G.J. is indebted to Dr. R. Buccheri of CNR, Palermo, Sicily for many
discussions on the importance of endophysics and to Jason Ridgway-Taylor for
discussions of Rota incidence algebras.

\end{document}